\documentclass[showpacs,APS]{revtex4}

\usepackage{graphicx}
\DeclareGraphicsExtensions{.jpg,.pdf,.png,.tif,.mps}
\usepackage{amsmath,amssymb}
\usepackage{color}
\usepackage[normalem]{ulem}
\usepackage{multirow}

\DeclareRobustCommand\bblash{\btt{\@backslashchar}} \makeatother

\begin{document}
\title{Pure Lovelock Kasner metrics}
\author{Xi\'an O. Camanho }
\affiliation{Max-Planck-Institut f\"ur Gravitationsphysik, Albert-Einstein-Institut, 14476 Golm, Germany}
\author{Naresh Dadhich}
\affiliation{Centre for Theoretical Physics, Jamia Millia Islamia, New Delhi 110 025, India, and  Inter-University Centre for Astronomy \& Astrophysics,\\ Post Bag 4 Pune 411 007, India}
\author{Alfred Molina}
\affiliation{Departament de F\'{\i}sica Fonamental, Institut de Ci\`encies del Cosmos. Universitat de Barcelona, Spain}
\date{\today}

\begin{abstract}
We study pure Lovelock vacuum and perfect fluid equations for Kasner-type metrics. These equations correspond to a single $N$th order Lovelock term in the action in $d=2N+1,\,2N+2$ dimensions, and they capture the relevant gravitational dynamics when aproaching the big-bang singularity within the Lovelock family of theories. Pure Lovelock gravity also bears out the general feature that vacuum in the critical odd dimension, $d=2N+1$, is {\it kinematic}; {\it i.e.} we may define an analogue Lovelock-Riemann tensor that vanishes in vacuum for $d=2N+1$, yet the Riemann curvature is non-zero. We completely classify isotropic and vacuum Kasner metrics for this class of theories in several isotropy types. The different families can be characterized by means of certain higher order 4th rank tensors. We also analyze in detail the space of vacuum solutions for five and six dimensional pure Gauss-Bonnet theory. It  possesses an interesting and illuminating geometric structure and symmetries that carry over to the general case. We also comment on a closely related family of exponential solutions and on the possibility of solutions with complex Kasner exponents. We show that the latter imply the existence of closed timelike curves in the geometry.
\end{abstract}

\maketitle

\section{Introduction}
Lovelock gravity is the most natural extension of general relativity (GR) in dimension higher than four, as it retains the basic character of the theory --the equation of motion remains second order despite the action being higher order in Riemann curvature. No other purely gravitational theory preserves this crucial feature. Yet another interesting feature of GR is the fact that it is {\it kinematic} in three dimensions and it turns dynamical in the next even dimension, {\it i.e.} $d=4$. In three dimensions the Riemann curvature tensor can be written in terms of the Ricci so that there exist no non-trivial vacuum solution. If we want this property to remain true as we go to higher odd dimensions there is a unique choice that corresponds to pure Lovelock gravity \cite{dgj, xd}. The action reduces in this case to a single $N$th order Lovelock term, with or without cosmological constant \footnote{The vanishing cosmological constant case, that will be the focus of this paper, has been also referred to as Chern-Simons or Born-Infeld gravity in odd and even dimensions respectively (see for instance \cite{Zanelli2000a}).} in dimensions $d=2N+1,2N+2$. This is the maximal order term in the Lovelock series, higher order terms being either topological or zero. There is no sum over lower order terms and in particular there is no Einstein term in the action. 

It is possible to define an analogue of the Riemann tensor for $N$th order Lovelock gravity, its characterizing property being that the trace of its Bianchi derivative vanishes, yielding the corresponding divergence free analogue of the Einstein tensor \cite{bianchi}. This is exactly the same as the one obtained from the variation of the $N$th order Lovelock action term. For the appropriate definition of Lovelock-Riemann tensor and zero cosmological constant, any pure Lovelock vacuum in odd $d = 2N + 1$ dimensions is Lovelock flat, {\it i.e.} any vacuum solution of the theory has vanishing Lovelock-Riemann tensor \cite{xd, kastor}.  Likewise, even for non-zero cosmological constant, the Weyl curvature vanishes in $3$ dimensions, and so does the Lovelock-Weyl tensor for pure Lovelock in all odd $d=2N+1$ dimensions. That is, pure  Lovelock gravity is {\it kinematic} relative to the Lovelock analogue of the Riemann tensor in all odd $d=2N+1$ dimensions. 

Kinematicity of pure Lovelock gravity was shown to be true for static spacetimes \cite{dgj} and conjectured to hold in general. This has indeed been proven in general recently \cite{kastor,xd} using a different definition for the Lovelock-Riemann tensor due to Kastor \cite{kastor}. Using purely algebraic properties, it has been shown that this $N$th order Lovelock-Riemann tensor can be entirely written in terms of the corresponding Lovelock-Ricci in odd $d=2N+1$ dimensions. This clearly establishes kinematicity of pure Lovelock gravity, as defined above, in all generality. There is no other gravity theory satisfying this property. Furthermore, for kinematicity to hold in all odd dimensions we should restrict the pure Lovelock equation to two dimensionalities only, $d=2N+1, 2N+2$, else this property would be violated.  That is, for any given Lovelock order, $N$, the pertinent dimensions are only these two. We will be always referring to them when mentioning odd and even dimensions in the text. Conversely for a given dimension, $d$, the order is fixed as $N=[(d-1)/2]$. Henceforth by pure $N$th Lovelock we would refer to this theory in those only two relevant dimensionalities, $d=2N+1, 2N+2$. 

In this paper we focus on Kasner-type metrics in pure Lovelock gravity. These give a very interesting class of cosmological solutions, which are in fact the simplest instances of homogeneous but anisotropic spaces, and have been very effectively employed for studying the approach to the big-bang singularity \cite{Belinskii2006}. In the general relativistic case in four dimensions, the approach to the singularity oscillates between several Kasner-like phases, contracting along two axes and expanding along the third, followed by a switch over to a different expanding direction, and so on.  The subsequent phases and the transitions between them can be represented as a map from the space of Kasner solutions to itself, the {\it Kasner map}, that has a very nice geometric structure \footnote{A nice description can be found in \cite{Heinzle2009a}.}. In the higher dimensional setting, while  considering general Lovelock gravity, the leading order behavior close to the singularity is captured by the highest order term, thus pure Lovelock gravity would describe the relevant gravitational dynamics in the approach to the big-bang singularity within the full Lovelock family of theories.

The Kasner metrics were also very instrumental in finding the right definition of Lovelock-Riemann tensor verifying kinematicity \cite{xd}. For pure Lovelock static vacuum spacetimes in $d=2N+1$ both definitions of Lovelock-Riemann tensors due to Dadhich \cite{bianchi} and Kastor \cite{kastor} vanish and the difference became apparent only while studying solutions of reduced symmetry. In fact, pure Lovelock Kasner metrics in odd dimensions have zero Lovelock-Riemann in Kastor's formulation whereas Dadhich's analogue is in general non-zero. 

In the following sections we will analyze pure Lovelock equations for vacuum and perfect fluid spacetimes and obtain solutions in the Kasner class. In Sections II and III, we set up the Lovelock framework and write the equation of motion for Kasner spaces. Isotropy of spatial stresses is required for both vacuum and perfect fluid solutions. These conditions are solved in Section IV, finding several classes of solutions or isotropy types. Sections V and VI follow with a detailed analysis of perfect fluid and vacuum solutions respectively.  For each of the vacuum families of solutions, we compute higher order curvature tensors that will allow us to actually distinguish between the different families. Kastor's and Dadhich's Lovelock-Riemann tensors provide an efficient characterization of the solutions. Finally, in Section VII we consider a family of exponential solutions, very closely related to the Kasner class, and we conclude with a discussion. In the appendix we entertain the possibility of considering Kasner-type solutions with complex exponents. Even though in some cases the metric can still be brought to real form, these spaces contain closed timelike curves and thus cannot be considered as viable solutions of any gravitational theory.  

\section{Lovelock Lagrangian and equation of motion}
The action of Lovelock gravity and the corresponding equation of motion are given by a sum of homogeneous polynomial terms in the Riemann curvature, each of them multiplied by a coupling $c_k$ with length dimension $L^{2(k-1)}$ relative to the Einstein-Hilbert term. Action and equations for these theories are most simply written in terms of differential forms
\begin{eqnarray}
\mathcal{L}&=&\sum_{k=1}^Nc_k\,\frac{2^k}{(2k)!(d-2k)!}\epsilon_{a_1 a_2\cdots a_d}\,R^{a_1a_2}\wedge \cdots \wedge R^{a_{2k-1}a_{2k}}\wedge e^{a_{2k+1}}\wedge \cdots \wedge e^{a_d} ~,\\
G_{\ c}^{b}&=& \sum_{k=1}^N c_k\,\frac{2^{k-1}}{(2k)!(d-2k-1)!}\,\epsilon_{a_1 a_2\cdots a_{d-1} c}\,R^{a_1a_2}\wedge \cdots \wedge R^{a_{2k-1}a_{2k}}\wedge e^{a_{2k+1}}\wedge \cdots \wedge e^{a_{d-1}}\wedge e^b ~.
\label{einstein}
\end{eqnarray}
In this language, the torsion and curvature forms are defined via Cartan's structure equations,
\begin{eqnarray}
T^a &=& De^a=de^a+\omega^a_{\ b}\wedge e^b ~,\label{streq}\\
R^a_{\ b}&=&d\omega^a_{\ b}+\omega^a_{\ c}\wedge \omega^c_{\ b} ~,\nonumber
\end{eqnarray}
for which we have introduced a covariant exterior derivative, $D$, with the corresponding connection 1-form $\omega^a_{\ b}$, in addition to the usual exterior operator $d$. In order to make contact with the usual tensorial formulation one imposes that the torsion is zero and solves for the spin connection in terms of the vielbein. The equation of motion above is obtained upon variation with respect to the vielbein only, leaving the spin connection unchanged, as the spin connection variation is  proportional to the torsion, therefore set to zero.

Alternatively, one can introduce a set of $(2k,2k)$-rank tensors \cite{kastor} product of $k$ Riemann tensors, completely antisymmetric, both in its upper and lower indices,
\begin{equation}
\left.\right.^{(k)}\mathbb{R}^{b_1 b_2 \cdots b_{2k}}_{a_1 a_2 \cdots a_{2k}}= R^{[b_1 b_2}_{\quad \quad [a_1 a_2}\cdots R^{b_{2k-1} b_{2k}]}_{\qquad \qquad a_{2k-1} a_{2k}]}~.
\label{kastensor}
\end{equation}
With all indices lowered, this tensor is also symmetric under the exchange of both groups of indices, $a_i\leftrightarrow b_i$. In a similar way we will denote the contractions of $\mathbb{R}$ simply as
\begin{equation}
\left.\right.^{(k)}\!\mathbb{R}^{b_1 b_2 \cdots b_{J}}_{a_1 a_2 \cdots a_{J}}=\left.\right._{(k)}\!\mathbb{R}^{b_1 b_2 \cdots b_{J} c_{J+1} \cdots c_{2k}}_{a_1 a_2 \cdots a_{J} c_{J+1} \cdots c_{2k}} \qquad ; \quad \forall \, J<2k ~.
\end{equation}
In terms of these new objects we can now write
\begin{equation}
\mathcal{L}=-\sum_{k=1}^N c_k\, {}^{(k)}\mathbb{R} \qquad \text{and} \qquad
G^a_{\ b}=\sum_{k=1}^N c_k\left(k\,{}^{(k)}\mathbb{R}^a_{\ b}- \frac12 {}^{(k)}\mathbb{R}\,\delta^a_{\ b}\right)
\end{equation}
In any of the formulations, this reduces to the usual Einstein-Hilbert action ($N=1$) for $d=4$ with the first non-trivial $N=2$ Gauss-Bonnet (GB) correction appearing in five or higher dimensions. Note that the $N$th order term in the action gives a non-trivial contribution to the equations only for $d\geq2N+1$. The GB contribution to the action can be written
\begin{equation}
\mathcal{L}_{GB}\equiv R_{\mu\nu\rho\sigma}R^{\mu\nu\rho\sigma}-4R_{\mu\nu}R^{\mu\nu}+R^2
\end{equation}

Varying this term with respect to the metric tensor $g_{\mu\nu}$ (equivalently the vielbein), we obtain
\begin{eqnarray}
G^{(2)\mu}{}_\nu=2(R^\mu{}_{\sigma\rho\alpha}R_{\nu}{}^{\sigma\rho\alpha}
-2R^\mu{}_{\rho\nu\sigma}R^{\rho\sigma}
-2R^\mu{}_{\sigma}R^{\sigma}{}_\nu+RR^\mu{}_\nu)-\frac{1}{2}L_{GB}\delta^\mu{}_\nu ,\nonumber
\end{eqnarray}
the GB analogue of the Einstein tensor.

For reasons that will become clear later, we will be concerned here with the case of pure Lovelock theories. For any given dimensionality $d=2N+1$ or $d=2N+2$, we will consider the maximal degree $N$th order term as the only one in the Lovelock series, {\it i.e.} $c_N=1$ and $c_k=0$ for $k\neq N$.

This particular class of theories has some special properties not shared by any other Lovelock theory. Notice that it includes Einstein gravity in three and four dimensions. In fact, pure Lovelock theories will preserve some features of $d=3,4$ general relativity that are not respected by higher dimensional Einstein's theory \cite{dgj,dgj2,dad2}. As already mentioned, in dimension three the Riemann tensor can be written in terms of the Ricci, in such a way that Ricci flat solutions are actually completely flat, they have zero Riemann. In dimension four this does not happen and the Weyl tensor is in general non-zero. Pure Lovelock gravities as described above generalize this property to any dimension, as it is easy to show. In odd $d=2N+1$ dimensions the ($2N,2N$)-rank tensor defined previously can be written completely in terms of the Lovelock-Ricci $\mathbb{R}^a_{\ b}$ (or the corresponding Einstein),
\begin{equation}
\mathbb{R}^{b_1\cdots b_{2N}}_{a_1\cdots a_{2N}}=\frac{1}{(2N)!}\epsilon^{b_1\cdots b_{2N+1}}\,\epsilon_{a_1\cdots a_{2N+1}}G^{a_{2N+1}}_{\qquad b_{2N+1}}~.
\label{main}
\end{equation}
which clearly shows that it vanishes when the corresponding Einstein tensor vanishes. However for even dimensionality, $d=2N+2$, the above expression is not valid. The analogous would not vanish because it expression involves the ($4$th rank) Lovelock-Riemann, $\mathbb{R}^{ab}_{\ \ cd}$. This proves in general, without any reference to any particular solution, that pure Lovelock gravity is {\it kinematic} in all odd $d=2N+1$ dimensions. In case of non-zero cosmological constant, it can be easily verified that the Lovelock-Weyl tensor, the traceless part of $\mathbb{R}^{ab}_{\ \ cd}$, vanishes as is the case for the usual Weyl in three dimensional Einstein gravity.

In all relevant odd or even dimensions the whole information about the $(2N,2N)$-rank tensor is contained in $\mathbb{R}^{ab}_{\ \ cd}$ so we will be referring to this quantity only for all practical purposes. We will be comparing the results obtained for this Lovelock-Riemann tensor with those of other 4th rank tensors defined below. 

In particular for $N=2$ we may define two $4$th rank tensors (see \cite{xd} for details and definitions for any $N$)
\begin{equation}{}^{(2)}\mathcal{R}^{\alpha\beta}{}_{\chi\gamma}\equiv R^{\mu\nu\alpha\beta} R_{\mu\nu\chi\gamma}+ 4 R^{[\alpha\mu} R^{\beta]}{}_{\mu\chi\gamma}+R R^{\alpha\beta}{}_{\chi\gamma}
 \end{equation}
and
\begin{equation}
{}^{(2)}\mathcal{M}^{\alpha\beta}{}_{\chi\gamma}\equiv R^{[\alpha\mu}{}_{\chi\gamma} R^{\beta]}{}_\mu-R^{[\alpha}{}_{[\chi}R^{\beta]}{}_{\gamma]}+
R_{\mu[\chi\nu}{}^{[\alpha}R^{\beta]\mu}{}_{\gamma]}{}^\nu
\end{equation}
Related to the first of these objects, one of us \cite{bianchi} defined an alternative Lovelock-Riemann analogue

\begin{equation}
{}^{(2)}\mathcal{F}^{\alpha\beta}{}_{\chi\gamma}\equiv {}^{(2)}\mathcal{R}^{\alpha\beta}{}_{\chi\gamma}-\frac{\mathcal{R}}{(d-1)(d-2)} \delta^\alpha{}_{[\chi}\delta^\beta_{}{\gamma]}
 \end{equation}
where $\mathcal{R}$ is the complete contraction of $\mathcal{R}^{\alpha\beta}{}_{\chi\gamma}$ and $d=5,6$. Kastor's Lovelock-Riemann \cite{kastor}, in turn, can be simply written as
\begin{equation}
{}^{(2)}\mathbb{R}^{\alpha\beta}{}_{\chi\gamma}\equiv \frac29\left(\frac14\, {}^{(2)}\mathcal{R}^{\alpha\beta}{}_{\chi\gamma} -  {}^{(2)}\mathcal{M}^{\alpha\beta}{}_{\chi\gamma}\right)
\end{equation}
In Sec. VI, it would be shown that for a particular class of pure GB vacuum solutions in odd $d=5$, ${}^{(2)}\mathcal{R}^{\alpha\beta}{}_{\chi\gamma}$ would be be non-zero while  ${}^{(2)}\mathbb{R}^{\alpha\beta}{}_{\chi\gamma}$ would be actually vanishing.

\section{Kasner metrics}

Consider a $d$-dimensional metric with flat $t=\text{const.}$ slices, for which each of the $n=d-1$ spatial directions scales differently with time in a polynomial fashion,
\begin{equation}
ds^2= \sum_{i=1}^n t^{2\,p_i} dx_i^2 -dt^2~.
\label{Kasner}
\end{equation}
This is known as the Kasner metric and is a homogeneous but anisotropic space. These type of metrics were also considered long ago by Deruelle \cite{Deruelle1989} in the GB case (see also \cite{Kitaura1991,Kitaura1993a} for a more general account of anisotropic models in Lovelock theories). 

For this class of metrics, each additional power of the Riemann curvature adds a factor $t^{-2}$ to the Lovelock-Riemann tensor and the strength of singularity increases accordingly with the Lovelock order.
Focusing in the dynamics near the big-bang singularity located at $t=0$, this means that the dominant contribution will come from the highest order Lovelock term in the action. Therefore, pure Lovelock theories capture the leading order behavior in this regime. 
In fact, Kasner metrics, as written above, will not, in general, be exact solutions of the equations of motion when considering terms of different curvature order in the action. If we want to analyze the subleading behaviour of the metric we may keep the next lower order Lovelock term and expand the metric further with two polynomial terms in time, $g_{ii}\sim t^{2p_i}+\beta_i t^{2q_i}$. Lower Lovelock terms become relevant for sufficiently long times, {\it i.e.} $\mathcal{L}_{N-k}/\mathcal{L}_N\sim c_{N-k}t^{2k}\sim 1$. 

Since for pure Lovelock the relevant dimensions are only two $d=2N+1, 2N+2$ we would therefore be considering the corresponding cases, $n=2N$ and $n=2N+1$,  respectively for odd and even dimensions. In what follows, we write the Lovelock-Einstein tensor, $G^{(N)}{}^a{}_b$, corresponding to the $N$th order Lovelock term, for the metric (\ref{Kasner}) for odd and even dimensions. Just the diagonal components are non-zero.

For odd $d=2N+1$ dimensionality, there are $2N$ exponents $p_i$ and the spatial components of the Lovelock-Einstein tensor are
\begin{equation}\label{eq14}
G^{(N)}{}^i{}_i=\frac{N(2N-2)!}{t^{2N}}\left( 2N-1-\sum_{m=1}^n p_m\right) \underbrace{p_j \cdots p_l}_{2N-1},\quad  m,j,l\neq i,\quad j<\cdots<l~,
\end{equation}
whereas the time component yields
\begin{equation}\label{eq15}
G^{(N)}{}^{t}{}_{t}=-\frac{N(2N-1)!}{t^{2N}}  p_1\cdots p_{2N}\,.
\end{equation}
Note that $G^{(N)}{}^i{}_i$ does not involve $p_i$; {\it i.e} there are only $2N-1$ exponents $p_{j\neq i}$, in the last product or the sum.

In particular, for $d=5$ GB these expressions reduce to
$$
G^{(2)}{}^1{}_1=\frac{4}{t^{4}}( 3-p_2-p_3-p_4)p_2p_3p_4
$$
and
$$G^{(2)}{}^{t}{}_{t}=-\frac{12}{t^{4}}  p_1p_2p_3p_4.$$

In even $d=2N+2$ dimensions the expressions are similar, yet a bit more involved,
\begin{equation}\label{eq16}
G^{(N)}{}^i{}_i=\frac{N(2N-2)!}{t^{2N}}\left( \left(2N-1-\sum_{m=1}^n p_m\right)\left(\sum_{j\cdots l=1}^{2N+1} \underbrace{p_j \cdots p_l}_{2N-1}\right)+ \underbrace{p_j \cdots p_l}_{2N}\right)\,\,j,k,l,m\neq i;\,\,j<k\cdots<l
\end{equation}
The number of $p_j$ in each product term in the sum is $2N-1$, and there are $\binom{2N}{2N-1}=2N$ such terms. Besides, the time component reads
\begin{equation}
G^{(N)}{}^{t}{}_{t}=-\frac{N(2N-1)!}{t^{2N}} \sum_{j,\cdots,l=1}^n \underbrace{p_j \cdots p_l}_{2N}=-\frac{N(2N-1)!}{t^{2N}}p_1\cdots p_{2N+1} \sum_{i=1}^{2N+1}\frac{1}{p_i},\quad j<k\cdots<l\label{tcomponent}
\end{equation}
for which each term in the sum is a product of $2N$ exponents $p_i$. Each product involves a combination of all the $2N+1$ $p_i$, {\it i.e} there are $\binom{2N+1}{2N}=2N+1$ terms.

For GB in $d=6$ we may write
$$
G^{(2)}{}^1{}_1=\frac{4}{t^{4}}\left[( 3-p_2-p_3-p_4-p_5)(p_2p_3p_4+p_2p_3p_5+p_2p_4p_5+p_3p_4p_5)+p_2p_3p_4p_5\right]
$$
and
$$G^{(2)}{}^{t}{}_{t}=-\frac{12}{t^{4}}(p_1p_2p_3p_4+p_1p_2p_3p_5+p_1p_2p_4p_5+p_1p_3p_4p_5+p_2p_3p_4p_5)=-\frac{12}{t^{4}}p_1p_2p_3p_4p_5\left(\frac1p_1\!+\!\frac1p_2\!+\!\frac1p_3\!+\!\frac1p_4\!+\!\frac1p_5\right) ~.
$$
\section{Isotropy conditions}

Even though Kasner spacetimes are anisotropic in general, in many situations we are interested in the isotropic case for which all the spatial components of the Lovelock-Einstein tensor are equal. This is the case for vacuum or perfect fluid solutions that will be the focus of our analysis in the following sections. We shall therefore analyze the isotropy conditions
\begin{equation}
G^{(N)}{}^i{}_i-G^{(N)}{}^j{}_j=0
\label{isocond}
\end{equation}
for every pair $i,j$. Not all equations written in this way are independent, only $n-1$ are, and we can choose, for instance, a set with fixed $i$ or $j$, say $i=1$.

\subsection{Odd dimension $d=2N+1$}

For odd dimensionality these conditions read
\begin{equation}
G^{(N)}{}^i{}_i-G^{(N)}{}^j{}_j\sim(p_i-p_j)\left(\sum_{k=1}^{2N}p_k-2N+1\right) \underbrace{p_l\cdots p_m}_{2N-2}=0, \quad l\cdots m \neq i,j.
\end{equation}

In particular fr five dimensional pure GB, it would reduce to
\begin{equation}
G^{(2)}{}^1{}_1-G^{(2)}{}^2{}_2\sim(p_1-p_2)\left(p_1+p_2+p_3+p_4-3\right) p_3 p_4 =0
\end{equation}
and two similar equations for $G^{(2)}{}^1{}_1-G^{(2)}{}^3{}_3$ and $G^{(2)}{}^1{}_1-G^{(2)}{}^4{}_4$.

\subsection{Even dimension $d=2N+2$}

In turn, in even dimensions we find
\begin{equation}
G^{(N)}{}^i{}_i-G^{(N)}{}^j{}_j\sim(p_i-p_j)\left(\sum_{k=1}^{2N+1}p_k-2N+1\right) \sum \underbrace{p_l\cdots p_m}_{2N-2}=0,\quad l\cdots m \neq i,j
\end{equation}
where each product in the sum contains $2N-2$ different $p_l$ that are a combination of all $(2N-1)$ available exponents with $l\neq i,j$, {\it i.e.} there are $\binom{2N-1}{2N-2}=2N-1$ of those terms.

in six dimensions, $N=2$, we shall write
\begin{equation}
G^{(2)}{}^1{}_1-G^{(2)}{}^2{}_2\sim(p_1-p_2)\left(p_1+p_2+p_3+p_4+p_5-3\right) (p_3 p_4+p_3p_5+p_4p_5) =0
\end{equation}
and three  similar equations for $G^{(2)}{}^1{}_1-G^{(2)}{}^3{}_3$, $G^{(2)}{}^1{}_1-G^{(2)}{}^4{}_4$ and $G^{(2)}{}^1{}_1-G^{(2)}{}^5{}_5\, .$ 

\subsection{Isotropy types}

There are several ways of solving the isotropy conditions. In both odd and even dimensions these reduce to a product of three factors. Therefore, for each pair $i,j$ at least one of those factors must vanish. Depending on the choice of the vanishing factors we will have several families of solutions or {\it isotropy types}.

\begin{itemize}

\item The first family --type {\bf (a)}-- corresponds to the trivial solution $p_i=p_j$ for every $i,j$, {\it i.e.} all the exponents are equal and the metric trivially isotropic.

\item Another very simple solution --type {\bf (b)}-- corresponds to putting to zero the second factor, that in fact does not depend on the pair $i,j$ that we pick,
\begin{equation}
2N-1-\sum_{k=1}^{d-1}p_k=0 ~.
\end{equation}
A single condition ensures isotropy  in this case. Whereas {\bf (a)} type solutions form a one parameter family of metrics, {\bf (b)} type spaces form a $n-1$ parameter family of solutions.

\item If we choose to set the third factor of the isotropy conditions to zero instead --type {\bf (c)}--, for every pair $i\neq j$ we get different conditions in odd and even dimensions. In odd $d=2N+1$ we have to solve
\begin{equation}
\underbrace{p_l\cdots p_m}_{2N-2}=0 \quad l\cdots m \neq i,j~,
\label{oddc}
\end{equation}
whereas in even $d=2N+2$ dimensions the condition is
\begin{equation}
\sum \underbrace{p_l\cdots p_m}_{2N-2}= \underbrace{p_l\cdots p_m}_{2N-1}\sum{\frac{1}{p_k}}=0 \quad l\cdots m,k \neq i,j~.
\label{evenc}
\end{equation}
The form of the solutions is similar though. As we will see below, in odd $d=2N+1$ this istropy type implies that two exponents vanish, say $p_1=p_2=0$, whereas in even $d=2N+2$ three exponents have to vanish, $p_1=p_2=p_3=0$ for instance. In both cases, these solutions correspond to vacuum spacetimes and will be considered in detail in Section VI. In the $N=1$ case with $d=3, 4$, $2N-2=0$, and hence these conditions become vacuous. There are no solutions in this class for Einstein-Hilbert gravity.

\item The last possibility --type {\bf (d)}-- is to combine isotropy types {\bf (a)} and {\bf (c)} in such a way that for some pairs $i,j$ we have $p_i=p_j$, whereas for others the third factor in the corresponding isotropy condition is the one that vanishes. As we will see below, this will only be possible in even $d=2N+2$ dimensions.

\end{itemize}

Isotropy types {\bf (a)} and {\bf (b)} are simple enough and do not require any more explanation. Let us discuss type {\bf (c)} and {\bf (d)} in more detail. In odd dimensions enforcing just one of type {\bf (c)} conditions readily implies that one of the exponents vanishes, say $p_1=0$. Automatically all the components of the Lovelock-Einstein tensor vanish except for $G^1_{\ 1}$, but isotropy implies that the remaining component must also vanish, and hence another $p_{i\neq 1}=0$. The rest of the exponents may take any value. To be more concrete let us focus for a moment in $d=5$ pure GB gravity, for which (\ref{oddc}) reds, say for $i=1$, $$ p_2p_3=0,\;p_2p_4=0,\; p_3p_4=0~.$$ This implies that two of the $p_i$ exponents with $i\neq 1$ must be zero. All the components of the Lovelock-Einstein tensor vanish in that case, therefore all solutions in this isotropy type are vacuum spacetimes. This is also true in general, that is, isotropy type {\bf (c)} in all odd $d=2N+1$ dimensions implies vacuum spacetime. We will comment more on this later on.

The even dimensional counterpart of the above statement is a bit more complicated to get. For $d=6$ pure GB again, we have for $i=1$,
$$  p_4p_5+p_3p_5+p_3p_4=0,\;p_4p_5+p_2p_4+p_2p_5=0,\;p_2p_3+p_2p_5+p_3p_5=0,\; p_3p_4+p_2p_4+p_2p_3=0$$
This implies that three of $p_i$ must be zero. Note that, if we set one of $p_i$ to zero, say, $p_1=0$, then the isotropy conditions would read as
$$  p_2(p_4p_5+p_3p_5+p_3p_4)=0,\; p_3(p_4p_5+p_2p_4+p_2p_5)=0,\;p_4(p_2p_3+p_2p_5+p_3p_5)=0,\;p_5(p_3p_4+p_2p_4+p_2p_3)=0~.$$
up to a factor $\sum p_i-2N+1$ that we asume to be non-zero. Otherwise the solution would already be considered in the type {\bf (b)} class. From the previous equations it is now obvious that two more $p_i$'s are zero. Again, this would lead to vacuum solutions which will be analyzed further in the corresponding section.

For general $N$, imposing the above condition (\ref{evenc}) just for one pair $i,j$ implies either two $p_{l\neq i,j}$ exponents are zero, for instance $p_1=p_2=0$, or else
$$ \sum_{k\neq i,j}\frac{1}{p_k}=0$$
In the first case, $p_1=p_2=0$, all components of $G^a_{\ b}$ but $G^1_{\ 1}=G^2_{\ 2}$ yield zero. Furthermore, isotropy requires the rest must also vanish which implies one more $p_{i\neq 1,2}=0$ and we have again a vacuum solution. The other possibility $\sum 1/p_{k\neq i,j}=0$, tacitly assumes that none of the exponents involved  $p_{k\neq i,j}$ is zero. However,  when considering another pair such that $p_l\neq p_j,p_i$, the only way $G^l_{\ l}-G^i_{\ i}=0$ is that at least one $p_{k\neq i,j,l}=0$, thus contradicting the initial assumption.

The only remaining possibility is that all $p_k$'s are organized in two groups such that either $p_k=p_{1,2}$ with $p_1\neq p_2$. This corresponds to isotropy type {\bf (d)} and we just have to impose one restriction
\begin{equation}
G^1_{\ 1}-G^2_{\ 2}\sim\sum_{k\neq 1,2}\frac{1}{p_k}=0~.
\end{equation}
The rest of the isotropy conditions are either equivalent to this one or trivial. For $p_{1,2}$
having multiplicities $n_{1,2}$ with $n_1+n_2=2N+1$, this reduces to
\begin{equation}
\frac{n_1-1}{p_1}+\frac{n_2-1}{p_2}=0
\label{typed}
\end{equation}
with $n_{1,2}\neq 1$. For $d=6$ GB gravity, we have for instance $n_1=3$, $n_2=2$ and $p_1=-2p_2$, and it is easy to check that the isotropy condition $G^1_{\ 1}=G^2_{\ 2}$ is verified.

\section{Perfect fluid}

Once the isotropy conditions are imposed the equation of motion coupled to matter has the form $T^\mu_{\ \nu}=\text{diag}(G^t_{\ t},G^i_{\ i},\ldots,G^i_{\ i})$, which matches precisely the form of the stress-energy tensor of a perfect fluid with energy density $\rho=-G^t_{\ t}$ and pressure $P=G^i_i$. Moreover, all components of the Lovelock-Einstein tensor $G^a_{\ b}$ scale as $t^{-2N}$ with time, and consequently the same behavior applies to the fluid $\rho$ and $P$ supporting these geometries. Energy and pressure are, in this way, proportional to each other and satisfy a linear barotropic equation of state of the form, $P=\omega\, \rho$ with $\omega=\omega(p_i;d,N)$.

\subsection*{{\bf (a)} \ All equal $p_i$ exponents}

We have density and pressure as given by

$$\rho= -G^{(N)}{}^{t}{}_{t}=\frac{(d-1)!}{2}\left(\frac{p}{t}\right)^{2N}$$
and
$$P=G^{(N)}{}^i{}_i=\frac{(d-1)!}{2}\left(\frac{p}{t}\right)^{2N}\left(\frac{2N}{d-1}\frac{1}{p}-1\right)$$
where $d=2N+1, 2N+2$ respectively for odd and even dimensions. The equation of state is
$$P=\left( \frac{2N}{d-1}\frac1p-1\right)\rho$$
which gives for $p=\frac{2N}{d-1}, \frac{3N}{2(d-1)}, \frac{N}{d-1}$ dust, radiation and stiff fluid respectively. It is an FRW flat model.

\subsection*{{\bf (b)} \ $\sum_{k=1}^{d-1}p_i=2N-1$}

Note that the $G^{(N)}{}^i{}_i$ component is free of $p_i$. We can however substitute in Eqs (\ref{eq14}) and (\ref{eq16})
$
2N-1-\sum_{m\neq i} p_m=p_i
$
so that in both cases $G^{(N)}{}^i{}_i$ has the same form of $G^{(N)}{}^{t}{}_{t}$ except for a numerical factor. Thus, upon substitution, all spatial stresses become equal, and it turns out that
$$P=\frac{1}{2N-1}\rho$$
for both odd and even dimensions. 

\subsection*{{\bf (c)} \ $d-2N+1$ zero exponents.}

As explained before this isotropy type corresponds to two and three zero exponents in odd and even dimensions respectively. This, as already mentioned, are vacuum solutions and thus have vanishing energy density and pressure, $\rho=P=0$.

\subsection*{{\bf (d)} \ $p_i=p_{1,2}$ with multiplicities $n_{1,2}$}
 For the isotropy condition {\bf (c)}, this is the only case of non-vacuum solution and this class of perfect fluid solutions exist only for even $d=2N+2\geq6$ dimensions. As before we have to further impose the condition,

$$ \frac{n_1-1}{p_1}+\frac{n_2-1}{p_2}=0$$ with $n_{1,2}\neq 1$ and $n_1+n_2=2N+1$.
Substituting this into the equations of motion we get
\begin{equation}
G^t_{\ t}\sim -(2N-1)p_1^{n_1-1}p_2^{n_2-1}(p_1+p_2)
\end{equation}
whereas the spatial components are
\begin{equation}
G^1_{\ 1}=G^2_{\ 2}\sim p_1^{n_1-1}p_2^{n_2-1}\left(2N-1-(n_1-1)p_1-(n_2-1)p_2\right)
\end{equation}
Thus we get an equation of state,
$$ P=\left(\frac{2N-1-(n_1-1)p_1-(n_2-1)p_2}{(2N-1)(p_1+p_2)}\right)\rho =\frac{1-\Delta\chi}{\Delta\chi}\rho~.$$
The nice parametrization above corresponds to $p_1=(n_1-1)\chi$, $p_2=-(n_2-1)\chi$ and $\Delta=n_1-n_2$. Notice that this class of solutions allows very large values for  $p_{1,2}$ (equivalently $\chi$) in which case we obtain $P=-\rho$. Conversely for small $p_{1,2}\rightarrow 0$ we have an {\it almost vacuum} solution with $P\approx (\frac{\rho}{p_1+p_2}$). The parameter $\omega=P/\rho$ thus spans the whole range $[-1,\infty)$ in this case.  In principle this parametrization would allow for fluids with lower barotropic parameter $\omega$ but these would violate all of the usual energy conditions.
Notice that the barotropic constant depends on two parameters in this case, one discrete, $n_1$, and the other continuous, $p_1$. For each possible value of $1<n_1<2N$ (or $\Delta$) we have a continuous function, $\omega=\omega(p_1)$.

\section{Vacuum solutions}
Vacuum spacetimes can be considered as a subset of the more general perfect fluid class of solutions from previous section. Sincein all cases we have a equation of the type $P=\omega \rho$, we just need to impose one more condition, $\rho=0$, to get the vacuum solutions. Clearly a perfect fluid or vacuum solution for pure Lovelock would not in general satisfy the isotropy condition for the Einstein tensor. 

In odd $d=2N+1$ dimensions $\rho=0$ implies that at least one of the exponents has to be zero. This fact combined with the isotropy conditions yields two types of solutions. Considering the isotropy type {\bf (b)} we get solutions of the form
\begin{equation}
p_1=0\quad \text{and}\quad  p_2=2N-1-\sum_{i=3}^{2N}p_i~.
\end{equation}
We have already mentioned that all isotropy type {\bf (c)} solutions are vacuum. These have not one but two zero exponents, say
\begin{equation}
p_1=p_2=0.
\end{equation}
In accordance with the discussion above, these solutions will be referred to as type {\bf (b)} and {\bf (c)} vacuum metrics respectively. In the way defined so far, these two families have a non-empty intersection. To be precise we will consider as type {\bf (b)} metrics those with {\it only} one vanishing exponent, $p_{i>1}\neq 0$, in order to have mutually exclusive categories. 

These are all non-trivial vacuum Kasner solutions of the theory in odd dimension. Type {\bf (a)} solutions are trivial as all the exponents are equal and thus $\rho=0$ implies that the metric is just Minkowski. Moreover, there are no type {\bf (d)} vacuum metrics in odd dimensions. In fact, there exists no solution with all non-vanishing exponents $p_i$ in odd dimensions. Besides, in the type {\bf (b)} case at least one of the exponents has to be positive whereas the sign of the rest is a priori unconstrained. There are no further restrictions on the exponents, $p_{i>2}$, of type {\bf (c)} solutions, on their signs or otherwise. Both vacuum families of solutions have the same number of free parameters, we have $2N-2$ free exponents, $p_{i>2}$.

For even $d=2N+2$ dimensions, there are two ways of solving the vacuum condition $\rho=0$. We may have two flat directions, say $p_1=p_2=0$, or
$$\sum_{i=1}^{2N+1}\frac1{p_i}=0~.$$
This can be combined with the different isotropy types to give three different families of solutions. The first two families are analogous to those found in the odd dimensional case and correspond to having at least two zero exponents. For type {\bf (b)} metrics we have  \footnote{Remarkably one particular solution in both odd and even dimensions is that the nonzero $p$ are all equal to one. For $d=3, 4$ in Einstein gravity this is actually the only solution in this class.}
\begin{equation}
p_1=p_2=0\quad \text{and}\quad  p_3=2N-1-\sum_{i=4}^{2N}p_i
\end{equation}
whereas type {\bf (c)} solutions are all vacuum as explained before. These have three flat directions instead of just two,
\begin{equation}
p_1=p_2=p_3=0
\end{equation}
These two families of solutions will be referred to as vacuum types {\bf (b.1)} and {\bf (c)} respectively.  We again further constraint type {\bf (b.1)} metrics to have {\it only} two flat directions, {\it i.e.} $p_{i>2}\neq0$, so that there is no intersection between types {\bf (b.1)} and {\bf (c)}.

For the last family of solutions, corresponding to the case of all $p_i$ being non-zero --only compatible with type {\bf (b)}--, we have
\begin{equation}
\sum_{i=1}^{2N+1}p_i=2N-1 \quad \text{and} \quad \sum_{i=1}^{2N+1}\frac1{p_i}=0 ~.
\label{b2cond}
\end{equation}
This family will be referred as {\bf (b.2)} and is the familiar Kasner metric in its most generic form. Clearly, there have to be both positive and negative exponents $p_i$, at least one of each kind.  For type {\bf (b.1)}, as it happens in odd dimensions, at least one exponent has to be positive but there may be none of negative sign. There are no constraints on the signs for type {\bf (c)} metrics.

Again type {\bf (a)} solutions are trivial and correspond to Minkowski flat space and, even though we have type {\bf (d)} metrics in even dimensions, $\rho=0$ would require either $p_{1,2}=0$ or $p_1+p_2=0$, both conditions incompatible with Eq. (\ref{typed}). The latter condition would imply $n_1=n_2$, but this is impossible as the sum of the multiplicities has to be odd, $n_1+n_2=2N+1$. Therefore, there are no vacuum type {\bf (d)} solutions. All this discussion is summarized in  Table \ref{classes}.

In odd $d=2N+1$ dimensions we have two $(2N-2)$-parameter families of solutions whereas for even $d=2N+2$ dimensions we have three sectors, two $(2N-2)$ and one $(2N-1)$-parameter families of solutions. It is clear that for $d=3, 4$ type {\bf (c)} vacuum solutions are just the trivial flat solution while for $d>4$, we have non-trivial vacuum solutions in that class.

\begin{table}[ht]
\begin{tabular}{c|c||c|c||c}
type & $d$ & isotropy cond. & vacuum & vac. curvature\\
\hline\hline
{\bf (a)} & any $d$ & $p_i=p \ ,\quad \forall i=1,2\ldots d-1$ & $p=0$ & $R=\mathcal{R}=\mathbb{R}=0$\\
\hline
{\bf (b)} & $2N+1$ &  & $p_1=0$ & $\mathbb{R}=0~$ $^{(\#_1)}$\\
\hskip.4in {\bf --b.1--}\, & $2N+2$  & $\sum_{i=1}^{d-1}p_i=2N-1$ & $p_1=p_2=0$ & $\mathbb{R}=0~$ $^{(\#_1)}$\\
\hskip.4in {\bf --b.2--}\, & $2N+2$  &  & $\sum_{i=1}^{d-1}\frac1{p_i}=0$ & $ R,\mathcal{R},\mathbb{R}\neq 0$\\
\hline
{\bf (c)} & $2N+1$ & $p_1=p_2=0$ & \multirow{2}{*}{\it all} & \multirow{2}{*}{$\mathcal{R}=\mathbb{R}=0~$ $^{(\#_2)}$}\\
& $2N+2$ & $p_1=p_2=p_3=0$ & & \\
\hline
{\bf (d)} & \multirow{2}{*}{$2N+2$} & $p_i=p_{1,2}$ with multiplicities $n_{1,2}$ & \multirow{2}{*}{\it none} & \\
&  & $\frac{n_1-1}{p_1}+\frac{n_2-1}{p_2}=0~, \quad n_1+n_2=2N+1$ & &
\end{tabular}
\caption{Classification of isotropic and vacuum solutions.  ${(\#_1)}$ $\mathcal{R}=0$ as well for the subset of type {\bf(b)} ({\bf(b.1)} in $d=2N+2$) with all non-zero exponents equal to one, $p_{i\neq1,2}=1$. ${(\#_2)}$ For {\it flat Kasner} (naively type {\bf(c)}) we also have $R=0$. for $N=2$, except for these {\it exceptional} cases, all the tensors that are not in the table are non-vanishing.}
\label{classes}
\end{table}

\subsection{Curvature tensors}
We may now compute the different tensorial quantities defined in the Introduction, namely Kastor's $\mathbb{R}_{abcd}$ and the alternative $\mathcal{R}_{abcd}$, but also the Riemann tensor, $R_{abcd}$; for the different vacuum types. For simplicity we will denote these $4$th rank tensors just as $\mathbb{R}$, $\mathcal{R}$ and $R$ respectively, not to be confused with respective contractions. Note that Lovelock vacuum is defined by  $\mathbb{R}_{ab}=\mathcal{R}_{ab}=0$.

The computation of these tensors is very simple for type {\bf (c)} vacuum metrics for which both tensors are actually zero. This is easy to understand as in both tensors there is at least one set of antisymmetrized $2N$ indices. For type {\bf (c)} metrics, for both odd and even dimensions, we have at most $2N-2$ non-zero exponents, {\it i.e.} $2N-1$ non-flat directions including time. As a consequence, at least one of the $2N$ antisymmetrized indices will have to be on one of the flat directions. As any component of the Riemann tensor involving that direction is zero, for all dimensions and Lovelock orders we have $\mathcal{R}=0$ and ${}^{(2)}\mathbb{R}=0$ (also then $\mathcal{M}=0$). Another way of stating the same thing is that the spacetime has reduced effective dimension $2N-1$ (number of non-flat directions), with $d-2N+1$ flat directions added to it, and the corresponding Lovelock-Riemann is therefore zero. The Lovelock-Riemann tensor is non-trivial only in  $d_{\text{eff}}\geq 2N$. It would have to be $d_{\text{eff}}\geq 2N+1$ for the Lovelock-Einstein to be non-zero. Moreover, as discused in \cite{xd}, in $d_{\text{eff}}=2N$ the Lovelock-Riemann is completely determined by the Lovelock scalar, whereas for $d_{\text{eff}}=2N+1$ it can be given in terms of the Lovelock-Einstein.  Note that the Lovelock scalar is proportional to the trace of the Lovelock-Einstein except when $d_{\text{eff}}=2N$. In that case the Lovelock-Einstein vanishes but the corresponding scalar in general does not. It does vanish in vacuum as we will see below. Vacuum spacetimes have non-trivial Lovelock-Riemann only in dimension, $d_{\text{eff}}\geq 2N+1$, below that threshold all vacua are Lovelock-flat; {\it i.e.} the corresponding Lovelock Riemann is zero.

For type {\bf (b)} solutions in odd dimensions (or {\bf (b.1)} in even dimensions), a similar argument holds. The effective dimension is a priori $d_{\text{eff}}=2N$ in this case. In principle we have enough indices so that the antisymmetrization does not necessarily yield zero as before. We may realize however that, if all exponents are either zero or one, all components of the Riemann tensor with any time index vanish, $R^{0i}_{\ \ 0i}=p_i(p_i-1)t^{-2}=0$, thus effectively removing that direction as well. We reduce again $d_{\text{eff}}=2N-1$, thus implying $\mathcal{R}=\mathbb{R}=0$. Notice that all such solutions belong to types {\bf(b.1)} and {\bf (c)} (depending on the number of zeros and ones) in both the relevant odd and even dimensions. One can easily check that, at least for $d=5,6$ pure GB, type {\bf (c)} metrics, and those of type {\bf (b.1)} ({\bf (b)} in 5d) with all non-zero exponents equal to one, are the only solutions verifying $\mathcal{R}=0$.

Something similar happens for for the Lovelock-Riemann. In this case, as the effective dimension is $d_{\text{eff}}=2N$, this tensor can be written completely in terms of a scalar, basically the Lovelock term for that effective dimension \cite{xd}. Also the equation of motion will be given in terms of this invariant, therefore it has to vanish. We can explicitly check that by computing the effective Lovelock scalar,
\begin{equation}
\mathbb{R}_{\text{eff}}^{(2)}=\frac{(2N)!}{(2N-1)t^{2N}}\left[\sum_{i=1}^{2N-1}p_i -2N+1\right]\prod_{j=1}^{2N-1}p_j
\end{equation}
where we have set to zero, $p_{i>2N-1}=0$. For type ({\bf b}) metrics this quantity is zero, thus, the whole Lovelock-Riemann tensor is also zero. Note that the Lovelock-Enstein is not trivial in this case as the {\it real} dimension is bigger than $2N$. All of type {\bf (c)} and type {\bf (b.1)} metrics are Lovelock flat solutions. This had to be the case in odd $d=2N+1$ dimensions because of kinematicity, but it is a non-trivial statement in even dimensions. Contrary to $\mathcal{R}$ the Lovelock-Riemann tensor vanishes for all type {\bf (b.1)} solutions not just for those with $p_{i\neq1,2}=1$. 

Summarizing, the least restrictive condition among the ones we use is Lovelock flatness. The Lovelock-Riemann tensor, $\mathbb{R}=0$, singles out type {\bf (b.1)} (all type {\bf (b)} in odd dimensions) and {\bf (c)} metrics. $\mathcal{R}=0$ is verified just for all of type {\bf (c)} but just a subsector of type {\bf (b.1)} with all non-zero exponents equal to one. These spaces effectively have reduced effective dimensionality, $d_{\text{eff}}\leq2N-1$, thus any tensor with $2N$ antisymmetrized indices vanish for these metrics. 

Moreover, the Riemann tensor vanishes only when all exponents are zero except for one that may be zero or one. Both possibilities correspond to flat spacetime, the former being just Minkowski space in its canonical form. The latter --so called {\it flat Kasner}--, even though naively belonging to a different isotropy type, also corresponds to Minkowski space, a patch of it, in a different set of coordinates. Therefore the previous solutions are, in general Lovelock flat but not Riemann flat. Besides, only type {\bf (b.2)} vacuum solutions have a non-trivial Lovelock-Riemann tensor. One can in fact verify that, the Lovelock flatness condition for 6-dimensional pure GB metrics precisely implies that these have to have at least two flat directions, thus belong to types {\bf (b.1)} or {\bf (c)}. The other tensors are also non-vanishing for {\bf (b.2)} metrics.
We can use the previous results concerning the vanishing of the different curvature tensors to actually classify all vacuum Kasner solutions into the corresponding families. For that we just have to be careful in identifying the {\it exceptional} cases, {\it i.e.} type {\bf (b)} metrics with all non-zero exponents equal to one and {\it flat Kasner}.
\subsection{Structure of vacuum solution space: the {\it Kasner Shamrock}}
\begin{figure}
\begin{center}
\includegraphics[scale=1]{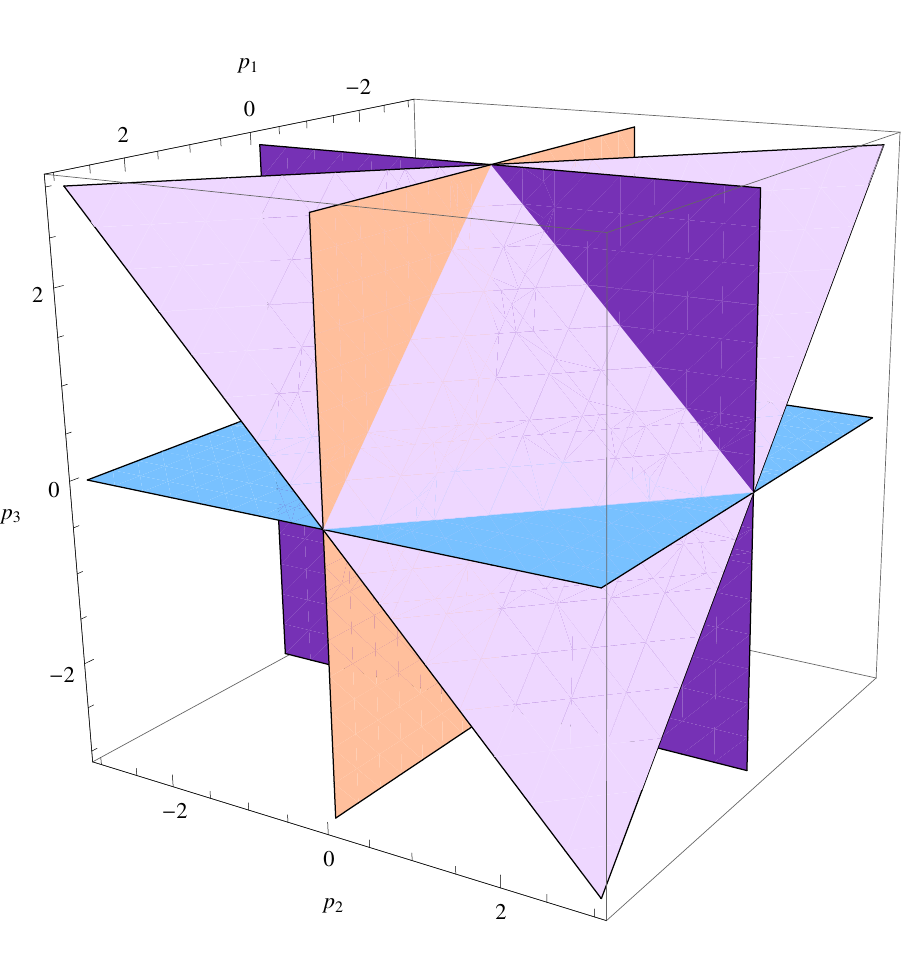}
\end{center}
\caption{Lovelock flat solutions in GB gravity in the $p_4=0$ plane in $d=5$ (also in $d=6$ with $p_5=0$). The vertical and horizontal planes correspond to each of the exponents being zero, $p_i=0$ for $i=1,2,3$, whereas the remaining one is $p_1+p_2+p_3=3$.}
\label{planes}
\end{figure}

The space of vacuum solutions can be easily visualized in the space parametrized by the exponents. Except for type {\bf (b.2)} that has a more complicated form, all of type {\bf (b)} and {\bf (c)} vacuum solutions correspond to intersections of planes in $\mathbb{R}^{d-1}$ (we will denote this space $\mathbb{R}^{d-1}_p$), two in odd $d=2N+1$ dimensions, and three in even $d=2N+2$ dimensions. These are the Lovelock flat solutions. For the simplest case, $N=2$, GB gravity in $d=5,6$ dimensions we can visualize this in the three dimensional parameter space  spanned by $(p_1,p_2,p_3)$ with the remaining exponents set to zero. We have in this way four families of solutions corresponding to  $p_1=0$, $p_2=0$, $p_3=0$ --type {\bf (c)}-- and $p_1+p_2+p_3=3$ --type {\bf (b)} or {\bf(b.1)} in odd and even dimensions respectively-- (see Figure \ref{planes}). 

The remaining case, type {\bf (b.2)}, has a much richer structure and will be explored in detail in what follows. These solutions verify the isotropy condition {\bf (b)}, $2N-1-\sum_{i=1}^{d-1} p_i=0$, which implies that the volume of $t=\text{const.}$  slices grows as $t^{2N-1}$. This is also true for vacuum type {\bf (b.1)} spaces. In addition, type {\bf (b.2)} solutions are the only ones that have a non-trivial Lovelock Riemann tensor, they are not Lovelock flat and are in this sense more {\it generic}. In fact they are in the most generic form of Kasner, with all non-vanishing exponents.

Type {\bf (b.2)} vacuum solutions verify one less condition than the other two  families present in even dimensions. In fact, we will see that type {\bf (b.1)} solutions appear at the boundary, in some appropriate sense, of the bigger {\bf (b.2)} family. When taking two of the exponents to zero, the second of (\ref{b2cond}) conditions have to be understood as a limit and, depending on how this is taken, we may solve the condition for any value of the remaining $p_i$. This condition is thus vacuous and we recover the {\bf (b.1)} type solutions.  These will appear as codimension one limiting surfaces in the space of {\bf (b.2)} solutions. Notice that type {\bf (c)} metrics are completely different in this sense, they cannot be recovered in this way as they correspond to different isotropy types. All type {\bf (b)} metrics lie in a plane $\sum p_i=2N-1$ in $\mathbb{R}^{d-1}_p$ whereas type {\bf (c)} metrics do not.

As we have two constraints on the exponents, we can always solve for two of them in terms of the remaining ones. Defining two new parameters
$$\zeta=p_1+p_2=2N-1-\sum_{k=3}^{2N} p_k\quad \mbox{and} \quad \xi=\frac1{p_1}+\frac1{p_2}=- \sum_{i=3}^{2N+1}\frac{1}{p_i}$$ we can determine $p_1$ and $p_2$ as solutions of a quadratic equation
$$p^2-\zeta\,p+\frac{\zeta}{\xi}=0$$
yielding
$$p_1=\frac{\zeta}2\left[1 \pm  \sqrt{1-\frac4{\zeta\xi}}\right],\quad p_2=\frac{\zeta}2\left[1 \mp \sqrt{1-\frac4{\zeta\xi}}\right]~,$$ all the remaining exponents $p_{k\neq 1,2}$ being free.  There is only one limitation to the above formula, the product of the two parameters cannot be in the $(0,4)$ range, therefore
\begin{equation}
\zeta\xi\geq 4 \quad \text{or} \quad \zeta\xi\leq 0~,
\label{reality}
\end{equation}
otherwise the values the $p_{1,2}$ become complex. This constrains possible values the remaining $p_i$ can take. For a pair of complex conjugate exponents we can make a complex change of variables making the metric real. These are also solutions to the pure Lovelock (also Einstein) equation, however they contain closed timelike curves and thus do not give physically interesting well defined spacetimes. We will comment more on this in the appendix \ref{app}.

Another interesting limiting case of the above formul\ae\ is $\zeta=0$ which necessarily implies $\xi=0$ as well, unless $p_1=p_2=0$ in which case we recover a type {\bf (b.1)} family of solutions. For $\zeta=\xi=0$, we can still choose the limiting value of $\zeta/\sqrt{\xi}$ in such a way that we obtain, $p_1=-p_2=p$, a one-parameter family of solutions. 

Alternatively, we may try to see what the space of allowed $p_3, p_4, \cdots$ looks like for fixed $\xi$ and $\zeta$, similar to how the Kasner solutions of Einstein gravity can be visualized as the intersection of a plane and a sphere
$$ \sum p_i=1\quad; \qquad \sum p_i^2=1. $$
In three dimensions these two constraints reduce the space of nontrivial solutions to just one, one of the $p_i$ parameters being zero and the other equal to one. This is what is known as the {\it flat Kasner} solution, as it has zero Riemann curvature. In four dimensions we can split the solutions into those with two vanishing exponents, the other being one --type {\bf (b.1)}, again {\it flat Kasner} in this case-- and those with all $p_i$ nonzero --type {\bf (b.2)}-- for which the sphere constraint is equivalent to the above $\sum 1/p_i=0$, with $i=1,2,3$, when the equation for the plane is used. The only type  {\bf (c)} metric is trivial in this case. The reality conditions (\ref{reality}) in this case simply amount to the exponents, all three, being in the range $p_i\in [-1/3,1]$. The upper bound is obvious from the sphere condition, otherwise at least one of the other exponents has to be complex.  One obvious difference between $d=4$ Einstein gravity and higher dimensional pure Lovelock is that the exponents are bounded as $\|p_i\|\leq1$ in the former case, equality meaning that the solution is {\it flat Kasner}. This seems to have some relation to the stability of these solutions \cite{Petersen2015}.

In $d=6$  GB gravity we can visualize the three dimensional space of $p_3,p_4,p_5$ in a similar way with equations,
$$ p_3+p_4+p_5=3-\zeta \quad ; \qquad \frac{1}{p_3}+\frac{1}{p_3}+\frac{1}{p_3}=-\xi. $$
For even dimensions higher than $6$ --and corresponding Lovelock order $N\geq 3$-- we can always decompose the space of exponents $p_i$ into lower dimensional subspaces similar to the one above. This will allow to suitably represent the space of solutions and also to exploit the rich symmetry structure of these equations. We now analyze the six dimensional case in detail. 

\begin{figure}
\begin{center}
\includegraphics[scale=.32]{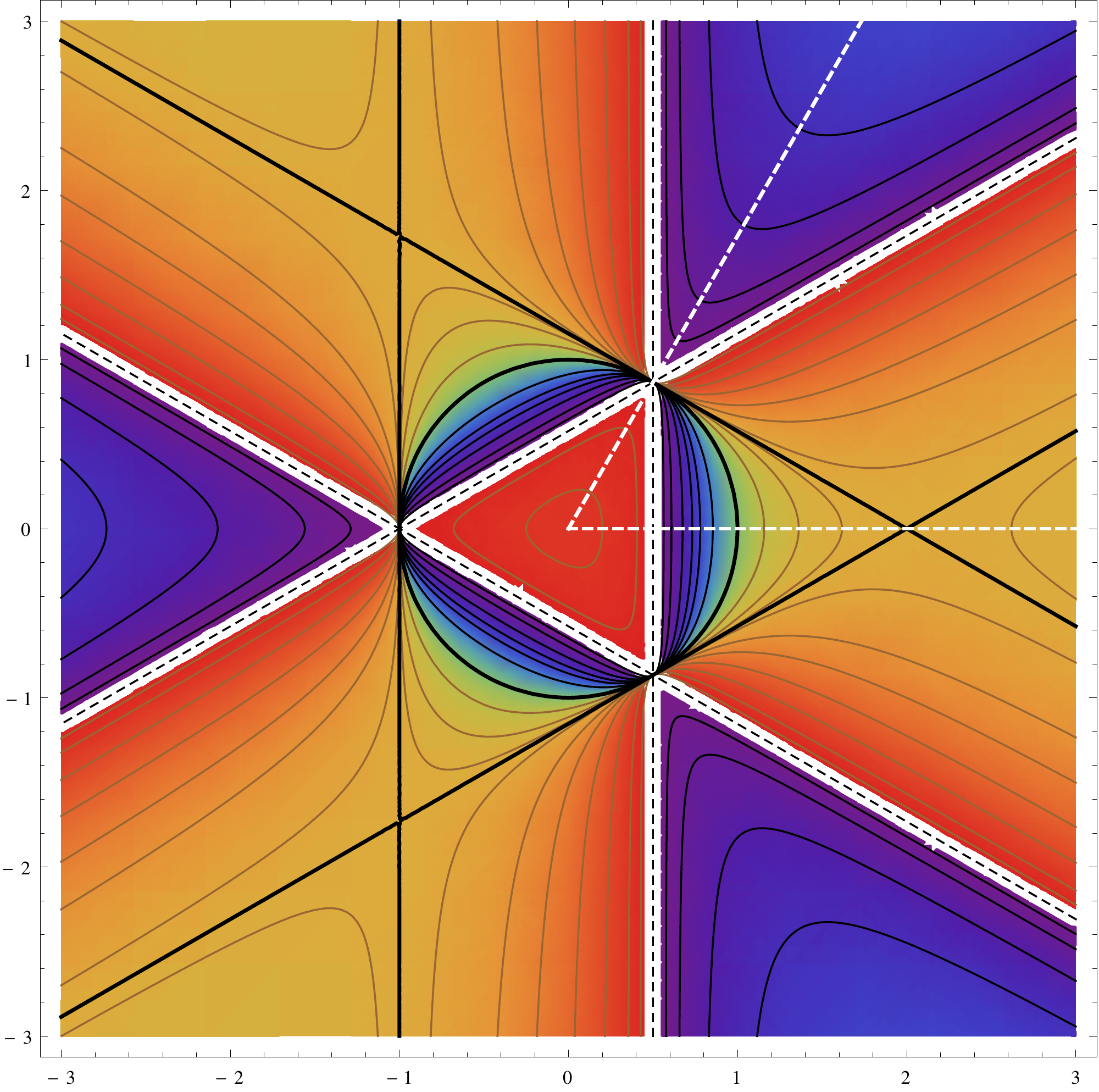}~
\includegraphics[scale=.32]{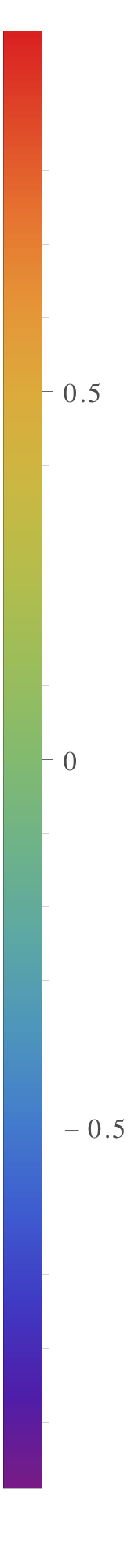}
\includegraphics[scale=.52]{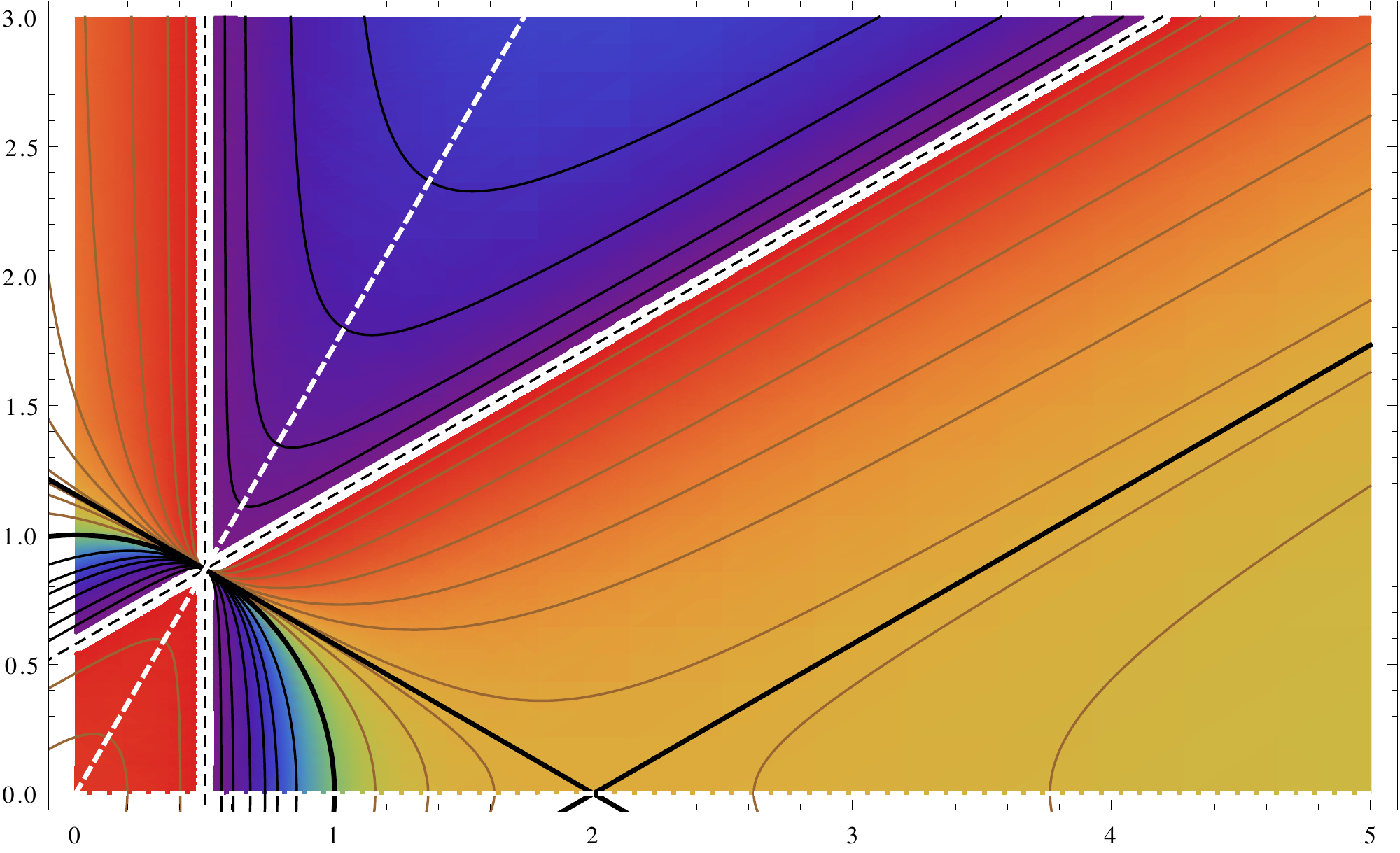}
\end{center}
\caption{Level diagram for the function $K(X,Y)=\sum_{i=1}^3 q_i^{-1}(X,Y)$ as defined in the text. The orbits of solutions for the different values of $k$ correspond to the level sets  of the function $K$, {\it i.e.} $K(X,Y)=k$. The numbers on the color scale correspond to $\frac{2}{\pi}\arctan(k)$.}
\label{Kspace}
\end{figure}

To make the analogy closer to the $d=4$ Einstein case we can rescale the exponents, $q_i=p_i/(3-\zeta)$, such that
$$ q_3+q_4+q_5=1 \quad ; \qquad \frac{1}{q_3}+\frac{1}{q_4}+\frac{1}{q_5}=\xi(\zeta-3) $$
The case $\zeta=3$ has to be treated separately. Notice that for $\xi=0$ we recover precisely the $d=4$ space of solutions, the {\it Kasner sphere} --a circle in this case--, whereas in general it gets deformed. To see how the deformation parameter, $$k=\xi(\zeta-3)=\left(\frac{1}{p_1}+\frac{1}{p_2}\right)\left(p_1+p_2-3\right)~ ,$$ affects the space of solution we can plot these spaces. In order to simplify the visualization we project onto the plane $q_3+q_4+q_5=1$ parametrizing $q_3=\frac{1+X-\sqrt{3}Y}{3}$, $q_4=\frac{1+X+\sqrt{3}Y}{3}$, $q_5=\frac{1-2X}{3}$ and plot in terms of the new $(X,Y)$ coordinates (see Figure \ref{Kspace}). Cold colors correspond to negative values of $k$ whereas warm colors indicate positive $k$. 
We see the thick line corresponding to the Kasner circle ($k=0$) and the white lines to one of the exponents being zero, $q_i=0$, or $k\rightarrow\pm\infty$.  These lines correspond precisely to type {\bf (b.1)} metrics. For $k$ to be infinite we need $\xi\to \pm\infty$ (we cannot take $\zeta\to \infty$ otherwise the rescaling would not be well defined) and thus either $p_1$ or $p_2$ vanish as well. The intersections of the white lines correspond to two $q_i=0$, the remaining {\bf(b.1)} solutions, for any value of $k$ in this case (equivalently any value of $p_{1,2}$). We can, in this way, analyze the whole set of type {\bf (b)} --both {\bf (b.1)} and {\bf(b.2)}-- vacuum even dimensional solutions. To complete the analysis one just has to include type {\bf (c)} solutions that cannot be represented in this way. 

Another interesting set of solutions is $k=1$, for which the orbit again yields three lines in the space of solutions. These correspond to each of the exponents being one, $q_i=1$ for $i=3,4,5$, that automatically solve the constraints. This same triangle plays a prominent role in $d=4$ Einstein gravity as the vertices can be used to define the {\it Kasner map}. This map is an application of the Kasner circle on itself and corresponds to the evolution of type II Bianchi models that have Kasner asymptotics both to the future and the past. The iteration of this map represents the so called {Mixmaxter attractor}, the subsequent transitions between Kasner epochs as we approach the big-bang singularity in more general (Bianchi IX) models (see \cite{Heinzle2009} for a recent discussion). It is reasonable to think that a similar map may exist for Lovelock gravity models as well,  despite the fact that the previous chaotic behaviour disappears in higher dimensions  for vacuum Einstein gravity \cite{Szydowski1987,Szydowski1987a,Turkowski1988}. The characterization of the space of solutions performed in this section can thus be taken as the starting point for the more general analysis of big-bang singularities for this class of theories. 

Interestingly enough, the previous representation makes apparent the  existence of three axes of symmetry that correspond to the exchange of a pair of exponents, $q_i\leftrightarrow q_j$. We have three possible such exchanges. We have also another symmetry that corresponds to a rotation of $2\pi/3$ that in turn represents a cyclic permutation of the three exponents. Using these symmetries we can always restrict the values of $X$ and $Y$ to one of the six fundamental domains, each corresponding to a different permutation of $(p_3,p_4,p_5)$.

The representation of Figure \ref{Kspace} is useful, for instance, to discuss the signs of the exponents. In the center of the inner triangle all three $q_i$ are positive, whereas each time one crosses one of the white lines one of the exponents changes sign. We can see thus that over the Kasner circle  we always have one negative and two positive exponents. This is also true for any solution corresponding to a positive value of $k$, except for the piece of the orbit inside the inner triangle (only for $k\geq 9$) that has all three exponents bigger than zero. Negative $k$ orbits split in two parts as well. The piece contained inside the Kasner circle $k=0$ still has two positive and one negative exponents, whereas the pieces in the outer blue triangles have two negative directions.

\begin{figure}[b]
\begin{center}
\includegraphics[scale=.7]{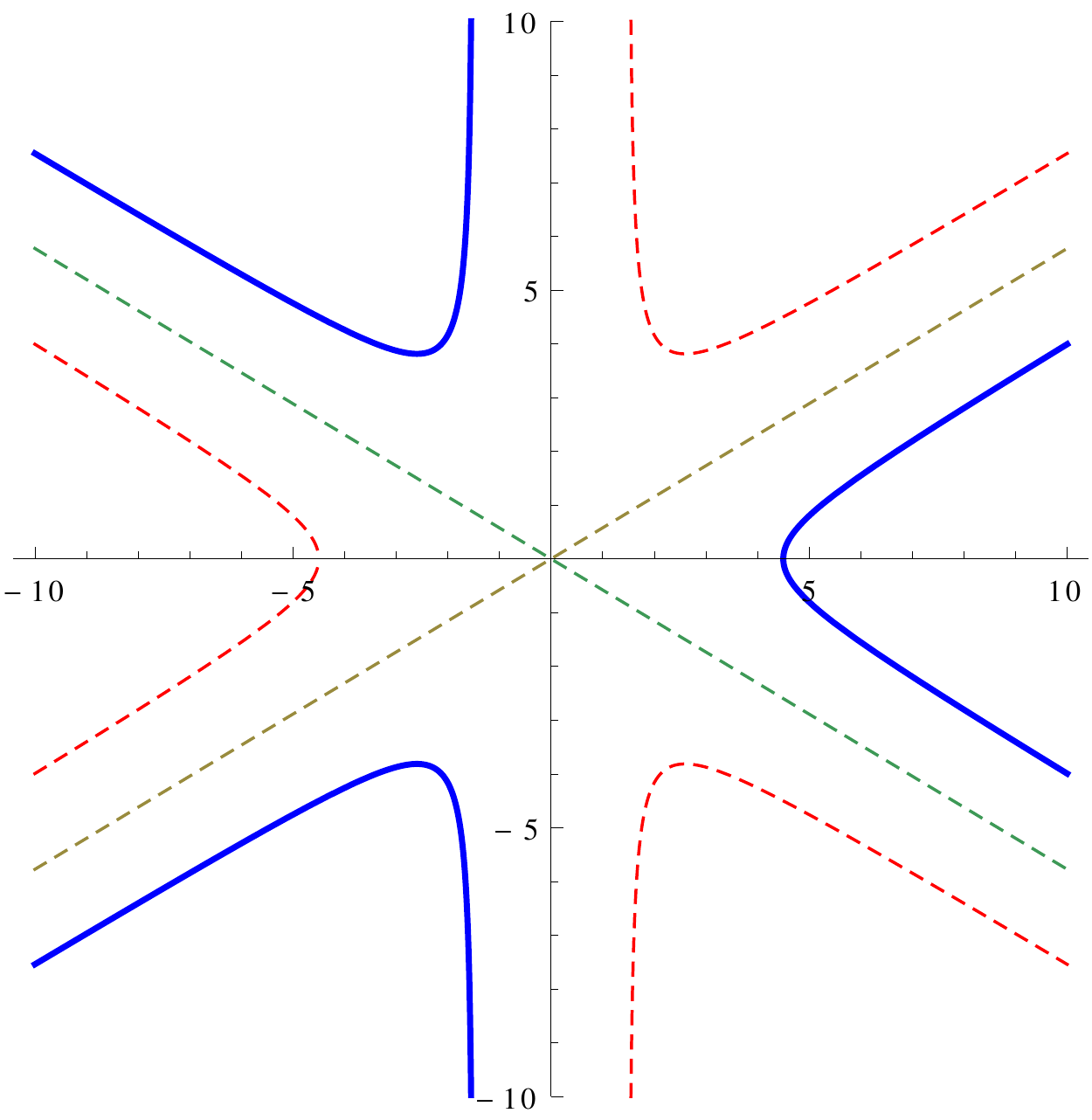}
\end{center}
\caption{Representation of the Kasner circle upon inversion as defined in Eq.~(\ref{invertedcircle}) with $\bar{q}_3=\frac{X-\sqrt{3}Y}{3}$, $\bar{q}_4=\frac{X+\sqrt{3}Y}{3}$, $\bar{q}_5=\frac{-2X}{3}$. }
\label{inversion}
\end{figure}

The only solutions that cannot be represented in the  $(X,Y)$ plane as above correspond precisely to $\zeta=3$. In this case the system reduces to
$$ p_3+p_4+p_5=0 \quad ; \qquad \frac{1}{p_3}+\frac{1}{p_4}+\frac{1}{p_5}=-\xi $$
and can be treated in a very similar manner. We can again rescale the exponents, $\bar{q}_i=-\xi \, p_i$, so that now we set to one the parameter in the second equation,
\begin{equation}
\bar{q}_3+\bar{q}_4+\bar{q}_5=0 \quad ; \qquad \frac{1}{\bar{q}_3}+\frac{1}{\bar{q}_4}+\frac{1}{\bar{q}_5}=1 
\label{invertedcircle}
\end{equation}
Using a similar projection as above,
$\bar{q}_3=\frac{X-\sqrt{3}Y}{3}$, $\bar{q}_4=\frac{X+\sqrt{3}Y}{3}$, $\bar{q}_5=\frac{-2X}{3}$
now on the plane $\sum_{i=3}^5 \bar{q}_i=0$, we can represent this orbit  as shown in Figure \ref{inversion} below (in blue).
Besides, if we change the signs of all the exponents $q_i$ we get the corresponding curves in the complementary regions (in red). Notice that the same system can be obtained from the generic case rescaling the $q$'s (or equivalently $X,Y$) by $k$ and taking the limit $k\rightarrow 0$.  These curves represent somehow the boundary of the $(X,Y)$ space above, brought to finite distance. Remark that in the original representation the $q_i=p_i/(3-\zeta)$ diverge as $\zeta\to 3$. Besides, we can also view this as a {\it Kasner sphere} with inverted coordinates, $q_i\rightarrow 1/\bar{q}_i$.

Instead of just for $\zeta=3$, we could have used the representation in terms of $\bar{q}_i$ and the corresponding rescaling from the beginning. We would then had gotten,
$$ \bar{q}_3+\bar{q}_4+\bar{q}_5=k \quad ; \qquad \frac{1}{\bar{q}_3}+\frac{1}{\bar{q}_4}+\frac{1}{\bar{q}_5}=1 $$
where the value of $k=\xi(\zeta-3)$ is the same as before.
In principle, any point can be represented in two equivalent ways related as $\bar{q}_i=k q_i$. In addition to this, we can also get one system of equations from the other through an inversion, $q_i\leftrightarrow 1/\bar{q}_i$. This inversion can be then rephrased as a symmetry of the original equations, whether in terms of $q_i$, $\bar{q}_i$ or $p_i$. This symmetry amounts to
\begin{equation}
\frac{p_i}{\zeta-3}\leftrightarrow\frac{1}{\xi p_i}\quad \text{or} \quad q_i\leftrightarrow \frac{1}{k q_i} \quad \text{or} \quad \bar{q}_i\leftrightarrow \frac{k}{\bar{q}_i}
\label{invsym}
\end{equation}
We can find similar symmetries for any subset of the exponents except for the complete set.

For any non-zero value of $k$, this symmetry would generate from one piece of the orbit some other piece with the same number of positive and negative exponents (for positive $k$) or all opposite signs (for negative $k$). The only points for which we cannot choose the representation we use are those with $k=0$. Depending on whether $\zeta=3$ or $\xi=0$, the corresponding orbit can just be plotted using $\bar{q}_i$ or $q_i$ respectively. Still we can easily obtain one of these orbits from the other through the inversion, $q_i\leftrightarrow 1/\bar{q}_i$.

The representation in terms of the $\bar{q}_i$ is otherwise not very useful.  The first representation, as already seen, is projected on the plane $q_3+q_4+q_5=1$ whereas for the second we can project on $\bar{q}_3+\bar{q}_4+\bar{q}_5=k$. Notice, that contrary to the original visualization scheme, now each point in the $(X,Y)$ plane does not have a unique value of $k$ associated to it, but three. There are three orbits with different values of $k$ through every point. It is much more complicated to plot and not very enlightening.

Remarkably enough, the inversion symmetry $q_i\leftrightarrow 1/(k\,q_i)$, when translated into the $(X,Y)$ plane, admits a very clear and elegant geometric realization. To see this, instead of performing just the inversion, we will define three different {\it inversion maps}, each of them being the composition of the inversion with one of the three possible exchange symmetries, $q_i\leftrightarrow q_j$ with $i\neq j$. Remember that these correspond to reflection symmetry about each of the three reflection axes of Figure \ref{Kspace}. We will refer to them as $f_i$ with $i=1,2,3$; {\it e.g.} $f_1$ corresponds to the composition of the inversion with the $q_2\leftrightarrow q_3$ exchange. The other maps are defined analogously.

\begin{figure}
\begin{center}
\begin{minipage}[b]{0.64\linewidth}
\includegraphics[scale=1.2]{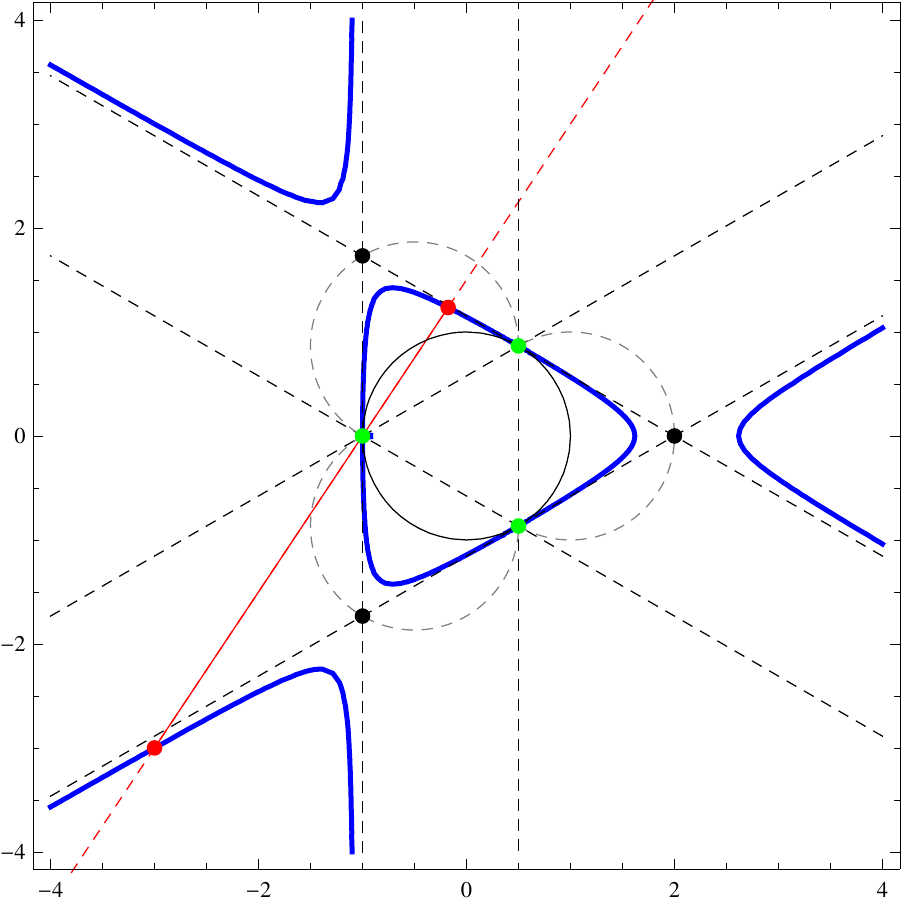}
\label{fig:minipage1}
\end{minipage}
\begin{minipage}[b]{0.34\linewidth}
\!\!\includegraphics[scale=.6]{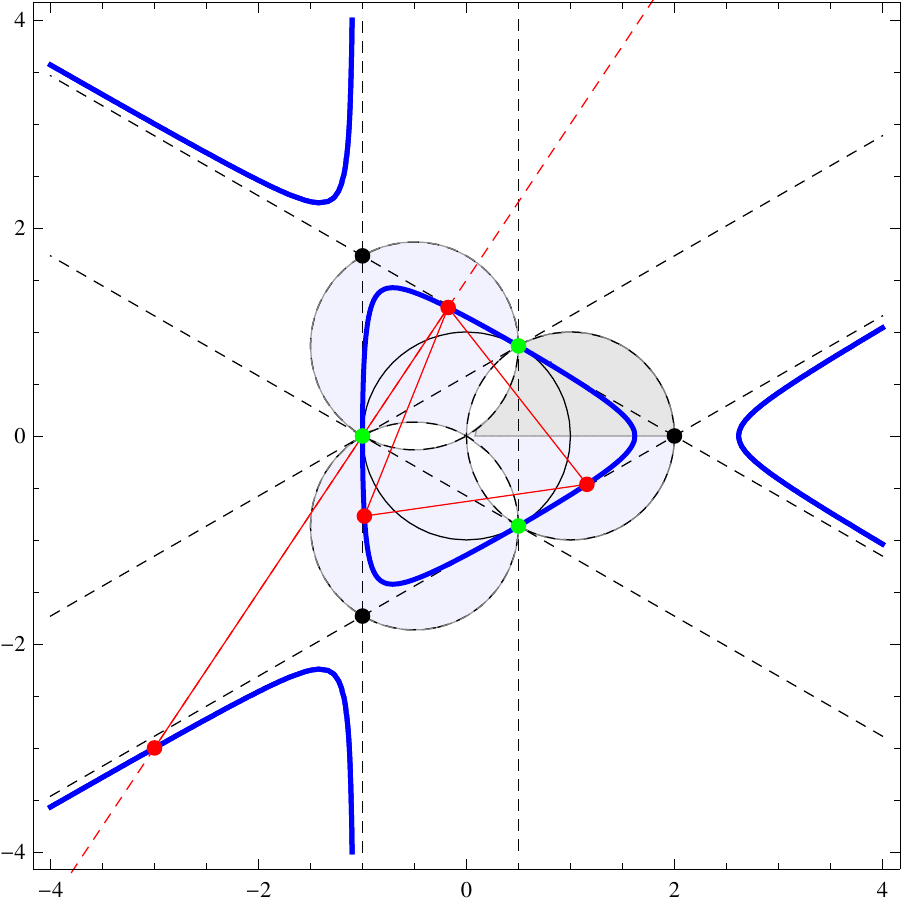}\\
\includegraphics[scale=.6]{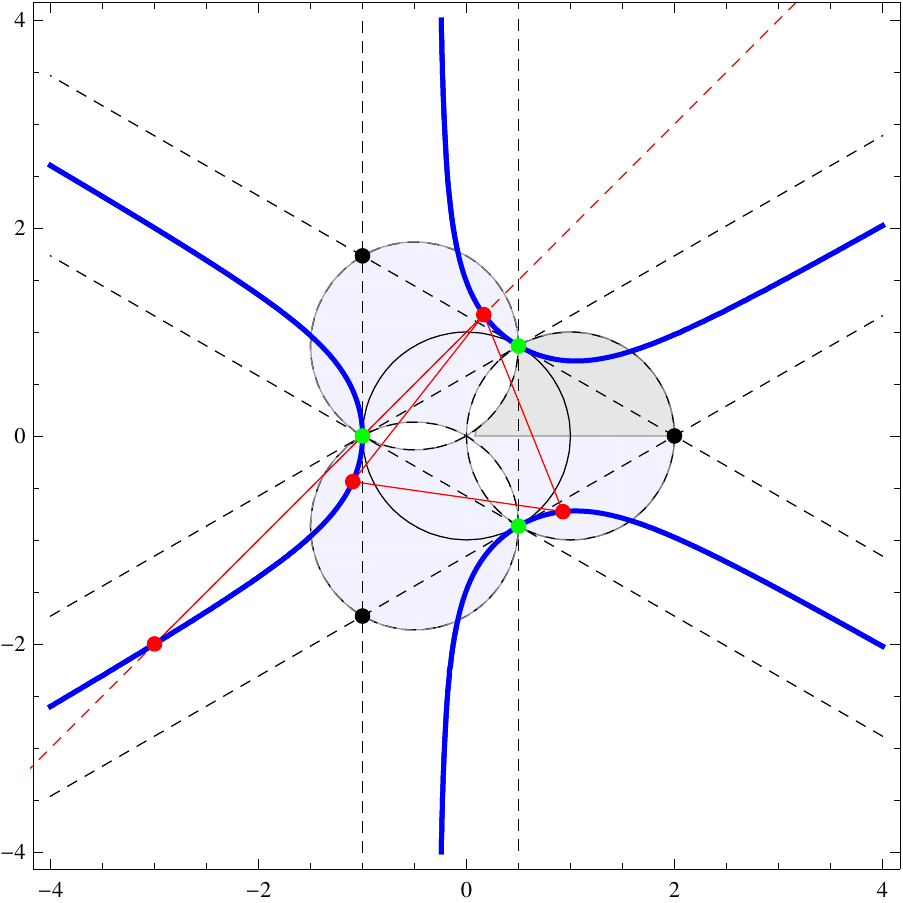}
\label{fig:minipage2}
\end{minipage}
\includegraphics[scale=.6]{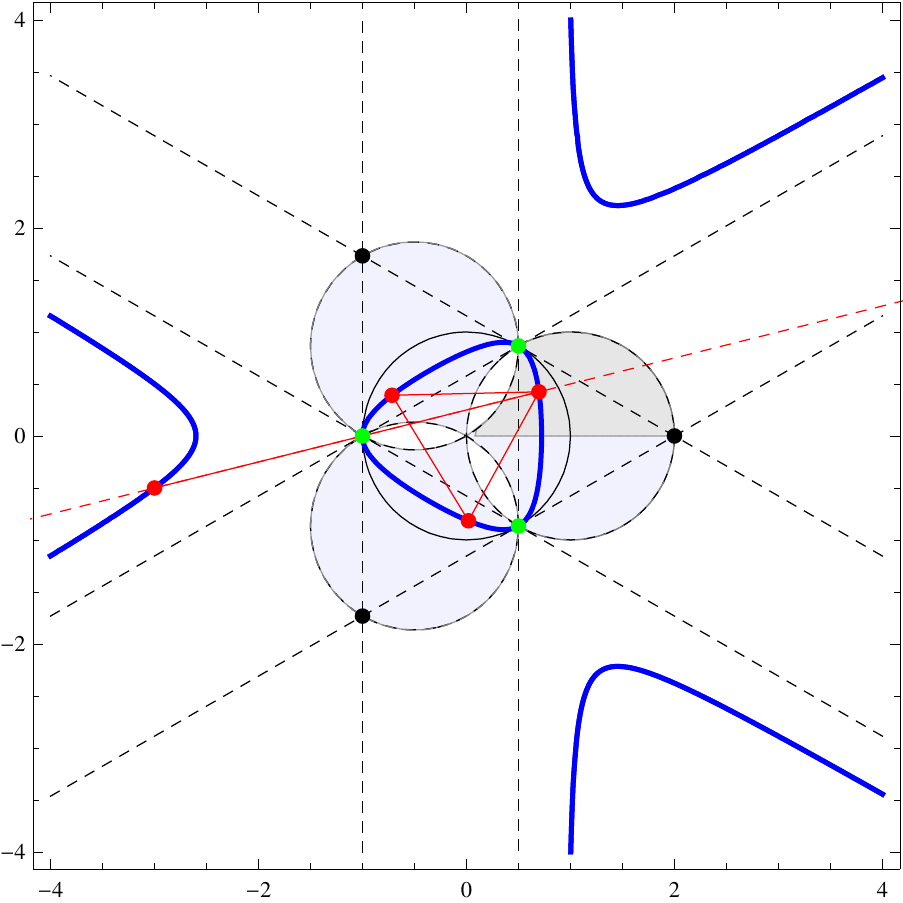}~~~\includegraphics[scale=.6]{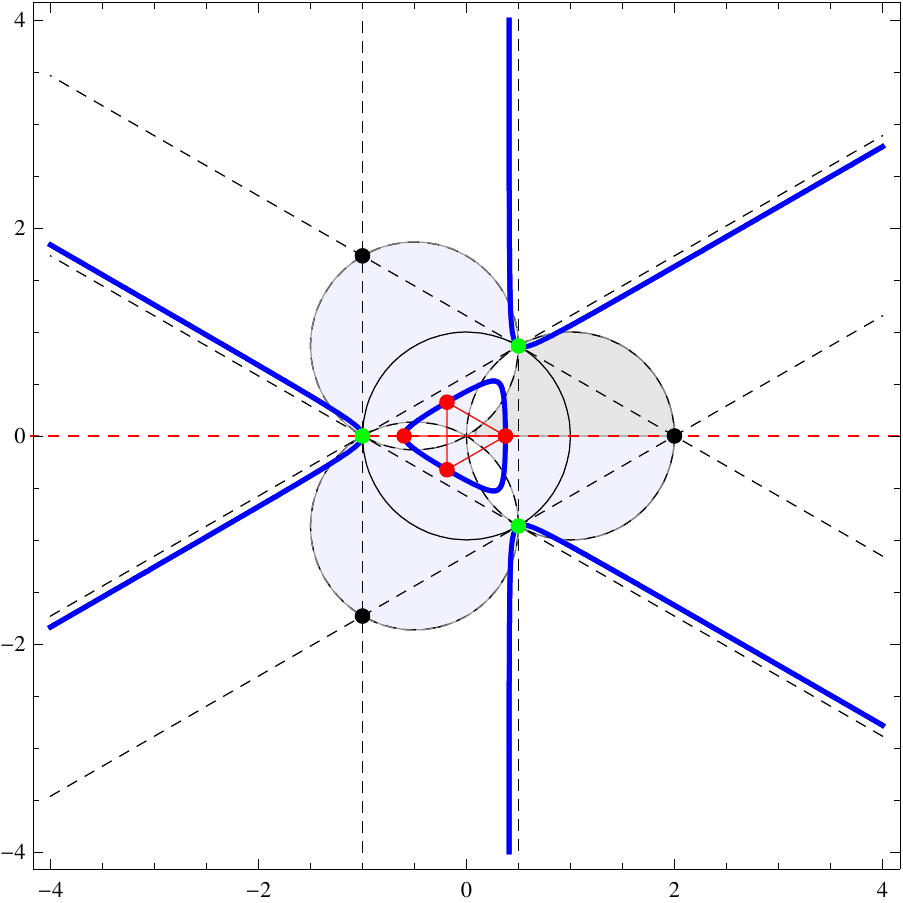}~~\includegraphics[scale=.6]{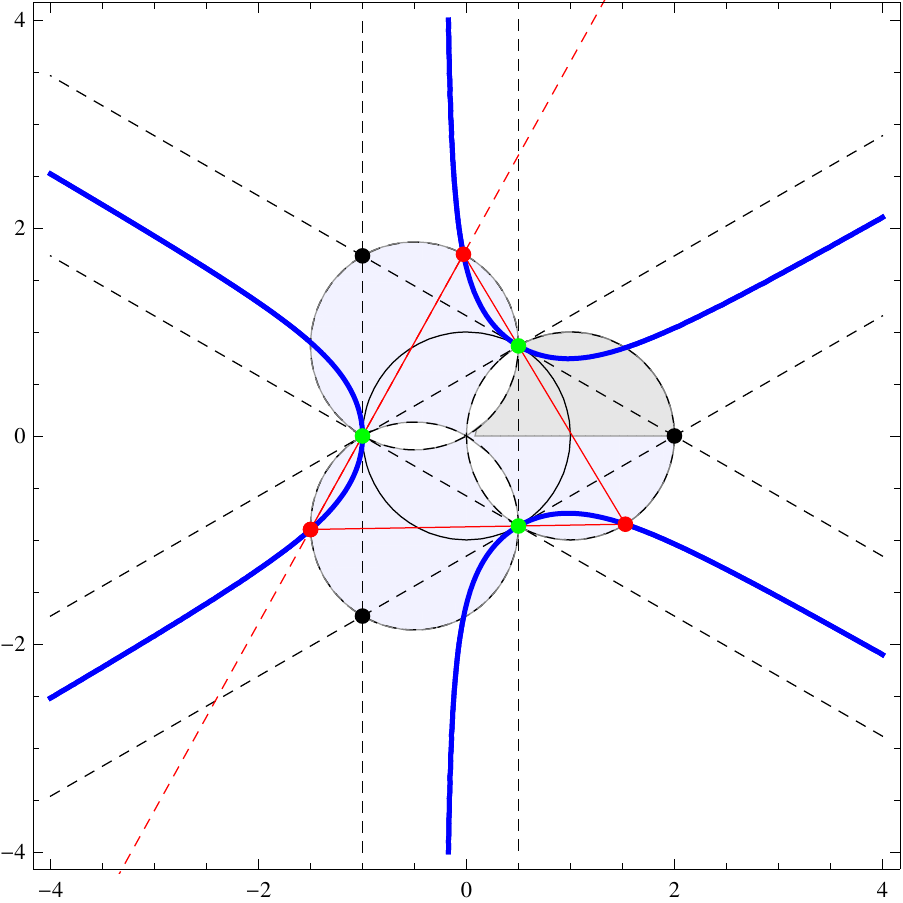}
\end{center}
\caption{Graphical representation of the inversion maps $f_i$. Any point and its image correspond to the intersection points of straight lines through each of the green points and a given $k$-orbit. The blue lines correspond to the orbits of solutions with the same value of $k$. For $f_3$ the related green point corresponds to $q_1=q_2=0$. An example of such map appears in the larger figure. The rest of the figures show all three images under the inversion maps for points in different regions of the solution space. The boundary of the {\it Kasner shamrock} corresponds to the points that are mapped to themselves under any of the inversion maps (see bottom right corner figure).}
\label{dualmap}
\end{figure}

The composition of these reflection symmetries with the inversion, each of these maps, verify a very interesting geometric property. The straight line on the $(X,Y)$ plane connecting the original point with its image via the map always passes through one of the intersections of two white lines, $q_i=0$ or $k\to\pm\infty$ (see Figure \ref{dualmap}, in green). Conversely, the intersection of one such straight line with the orbit of  given $k$ value would give me two points related by the map. This is very reminiscent of the way one can define geometrically the {\it Kasner map}, the only difference is the points used to trace the lines. The {\it Kasner map} has the intersections of the $k=1$ lines as focal points, instead of those of $k\to\pm\infty$. If I pick a different green point to trace the line the resulting points would be related through a different one of the maps. Moreover, from a given point, tracing the lines through the three green points I would get the corresponding three images. These images are all image under inversion of the same point, up to a reflection symmetry. Therefore they are related through the composition of two such reflections, {\it i.e.} they are related by a $2\pi/3$ rotation. Graphically they form an equilateral triangle around the origin of the $(X,Y)$ plane.

\begin{figure}
\begin{center}
\includegraphics[scale=.8]{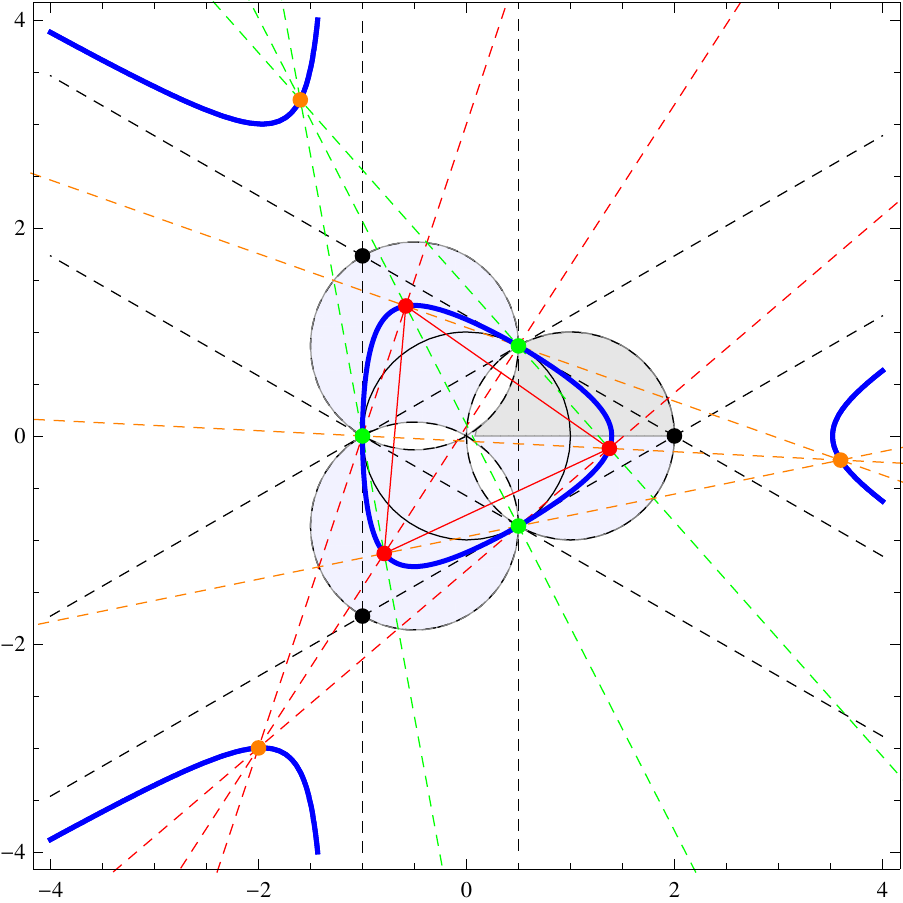}~~\includegraphics[scale=.8]{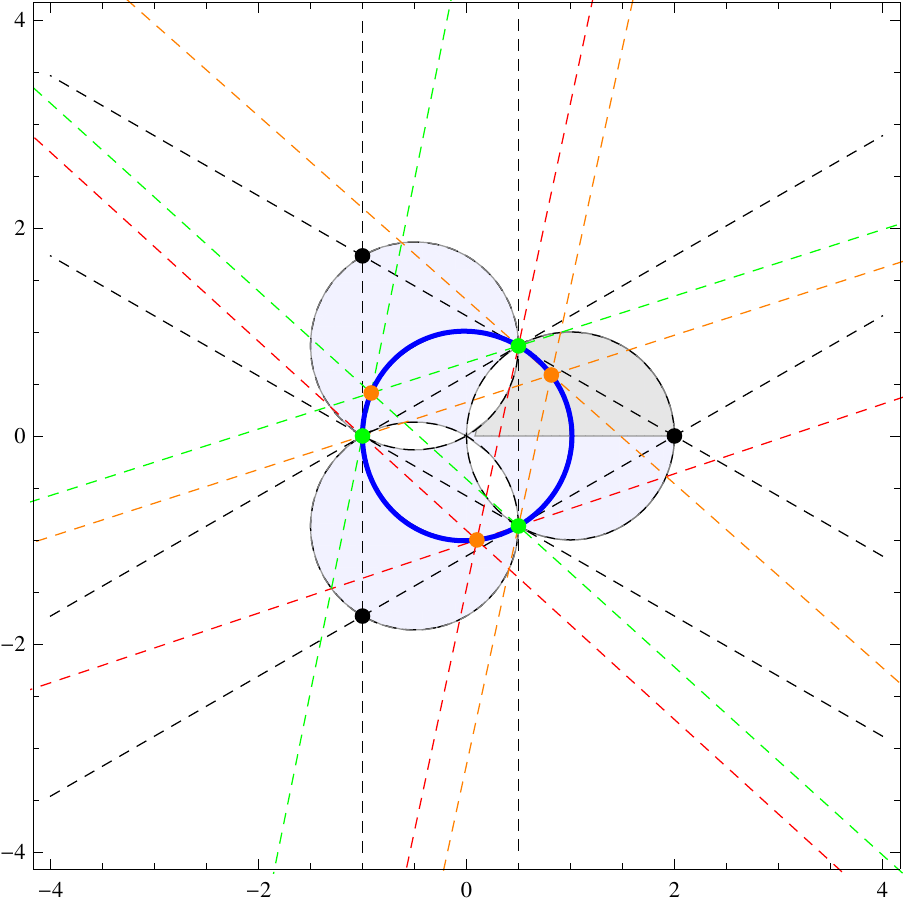}
\end{center}
\caption{Graphical representation of the inversion maps of three points related by $\pm 2\pi/3$ rotations in the $(X,Y)$ plane (on the left). The tracing of the nine straight lines joining these points (in red) with the points generating the maps (in green) determine the three corresponding image points (in orange). The only points that do not have an image in this way are those in the {\it Kasner circle} whose lines are parallel in groups of three (on the right).  }
\label{dualrev}
\end{figure}

Using this map we can restrict to the values of $(X,Y)$ that are inequivalent under the inversion. This will allow us to focus on a compact set of points in that plane, despite the original space of solutions being non-compact. The boundary of that compact set (shaded region in Figure \ref{dualmap}) is given by the fixed points of any of the maps. It is easy to verify that one such set of fixed points corresponds to the circle,
$$ (X-1)^2+Y^2=1 $$
whereas the rest can be obtained by rotating this set by $\pm 2\pi/3$ (see Figure \ref{dualmap}, bottom right). Each of the regions contained in the shaded region generates under inversion one of the corresponding regions outside. More precisely, given the set of solutions for a given $k$ inside the three circles we get its counterpart outside. We can actually perform this operation in a completely geometric way just tracing lines. Given three points related by $2\pi/3$ rotations inside the shaded region, one can trace the nine lines through these and the green points mentioned before. This nine lines will cross in groups of three outside the shaded region generating the images of the three original points under inversion (see Figure \ref{dualrev}). These are also related by $2\pi/3$ rotations. In this way, given the constant $k$ orbits in the shaded region we get the corresponding orbits outside.

This operation is well defined for all points on the $(X,Y)$ plane, except for the points on the {\it Kasner circle} that get mapped to infinity (the lines are parallel in groups of three), corresponding to the orbit $\zeta=3$, and points on the white lines. Each of this lines gets mapped to a single points, one of the green points used to generate the map. Notice that these are {\bf (b.1)} solutions that do not enjoy the inversion symmetry.

The more involved part of the shaded region corresponds to the inside of the inner triangle. In that region the circles overlap dividing the triangle in six regions, three {\it narrow} and three {\it wide} regions. The map thus maps the {\it narrow} regions into the {\it wide} ones and viceversa. We have chosen the {\it wide} regions to belong to our shaded {\it fundamental domain} but we could as well have chosen the other ones. The shape of this fundamental domain consists in three equal leaves of almost circular shape, joined at the center, reminiscent of a shamrock. We may call it the {\it Kasner shamrock}. Each of the leaves of the {\it shamrock} can be divided in two halves, each of these halves contained in one of the six sectors from which one can reconstruct the space of solutions with reflections. Thus, from just one of these half-leaves we can reconstruct the whole space of solutions using reflections and inversions.

\section{Exponential solutions}

Another class of solutions closely related to Kasner spacetimes is that of exponential type metrics of the form
\begin{equation}
ds^2=-dt^2+\sum_{i=1}^n e^{2H_it}dx_i^2
\end{equation}
These exponential solutions are a generalization of de Sitter spaces, with different {\it Hubble parameters} in every direction. This is very similar to the Kasner metric in a different form except that the $dt^2$ lacks the customary $e^{2t}$ in front. In fact, both Kasner and exponential solutions can be treated together using a slightly more general form of the metric,
\begin{equation}
ds^2=-e^{2H_0t}dt^2+\sum_{i=1}^n e^{2H_it}dx_i^2~,
\label{expansatz}
\end{equation}
that reduces to the Kasner metric for $H_0=1$ and to the exponential form for $H_0=0$. For more general values of $H_0$ we can always make a change of coordinates of the form $\tilde{t}=\frac1{H_0}e^{H_0t}$ (accompanied by a rescaling of the spatial coordinates) to bring the metric again to the Kasner form with exponents $p_i=H_i/H_0$. The exponential solutions can thus be pictured as living in the {\it boundary} of the space of Kasner solutions. We can take a limit where $p_i\to \infty$, at least for some of the Kasner exponents with, $p_i/p_j=H_i/H_j$.

 The role of the Hubble exponents, $H_i$, is very similar to the $p_i$ of previous sections. There are just two differences. First while the $p_i$ are dimensionless quantities the $H_i$ have dimension of inverse length. This is the reason the $(2,2)$ Riemann components scale as $t^{-2}$ for Kasner whereas they are constant in the exponential case. Second, the way these constants enter the Riemann curvature is just slightly different, the $R^{ti}_{\ \ ti}=\frac{p_i(p_i-1)}{t^2}\rightarrow H_i^2$ while the remaining components, $R^{ij}_{\ \ ij}=\frac{p_ip_j}{t^2}\rightarrow H_iH_j$, are equal in form without the $t^{-2}$ scaling factor.  For the more general ansatz the form of the $ti$ components would be $H_i(H_i-H_0)$ making the connection between the different parametrizations obvious. 

At the level of the pure Lovelock equations of motion, we realize that these become of homogeneous degree in the new exponents, $H_i$. From this basic observation and the discussion of Kasner solutions we can readily get the general form of the constraints for the exponential metrics. The algebraic form of the equations for the $H_i$ parameters is the same as for the $p_i$ when we keep just the highest degree in all expressions. Lower order in $p$ terms correspond to terms with $H_0$ factors in the exponential representation that are then set to zero. For instance we can take the above classification of vacuum solutions for pure Lovelock gravity and write: 

In odd $d=2N+1$ dimensions we have two $(2N-2)$-parameter families of solutions
\begin{itemize}
\item $H_1=H_2=0\ \ $ --type {\bf (c)}.
\item $H_1=0$ and $H_2=-\sum_{i=3}^{2N}H_i\ \ $ --type {\bf (b)}, with $H_{i>1}\neq0$, so that there is no intersection between types {\bf (c)} and {\bf (b)}.
\end{itemize}
whereas in even $d=2N+2$ dimensions we have three sectors, two $(2N-2)$ and one $(2N-1)$-parameter families of solutions,
\begin{itemize}
\item $H_1=H_2=H_3=0\ \ $ --type {\bf (c)}.
\item $H_1=H_2=0$ and $H_3=-\sum_{i=4}^{2N+1}H_i\ \ $ --type {\bf (b.1)}, again with $H_{i>2}\neq0$.
\item $H_i\neq 0, \quad \forall i$ with $\sum_{i=1}^{2N+1}H_i=0$ and $\sum_{i=1}^{2N+1}\frac1{H_i}=0\ \ $ --type {\bf (b.2)}.
\end{itemize}
This is summarized in Table \ref{classExp}. We can do the same for perfect fluid solutions or any other. Notice that this class of vacuum solutions does not exist in Einstein gravity without cosmological constant. Revisiting the {\it Kasner circle} adapted to this case we get
\begin{equation}
\sum_{i}H_i=0\qquad ; \quad \sum_i H_i^2=0
\end{equation}
and the second condition readily implies that all the exponents are zero.

In order to go to the boundary of the space of solutions, as explained above, we may take $p_{1,2}\to \infty$, therefore, either $\xi\to 0$ or $\zeta\to \infty$. In the latter case we can rescale $q_i=-p_i/\zeta$ ($p_i\to \infty$) to get the equation of the $k\to\pm\infty$ orbit
\begin{equation}
\sum_{i=3}^{d-1}q_i=1 \quad \text{and}\quad \sum_{i=3}^{d-1}\frac1{q_i}=\xi\zeta\to \pm\infty
\end{equation}
The second possibility is to rescale then $\bar{q}_i=-p_i\xi$ ($p_i\to \infty$) to get
\begin{equation}
\sum_{i=3}^{d-1}\bar{q}_i=\xi\zeta\to 0 \quad \text{and}\quad \sum_{i=3}^{d-1}\frac1{\bar{q}_i}=1
\end{equation}
This is the image of the Kasner circle under inversion $q_i\to 1/\bar{q}_i$ (see Figure \ref{inversion} for the $d=6$ case). Moreover, taking both limits at a time, $\xi\to 0$ and $\zeta\to\infty$ we can actually fix the value $\xi\zeta\to k$ getting the corresponding orbit, analogous to the Kasner case, and its image under inversion. The structure of the space of solutions is very similar to the Kasner case. Notice that the inversion $H_i\to 1/H_i$ is also a symmetry of the exponential space of solutions for the whole set of exponents $H_i$ in this case. For other subsets we can define an analogous transformation similar to that of previous sections. We can actually identify the parameters $q_i$ with a rescaled version of the Hubble parameters $H_i$.

We can readily analyze the form of the curvature 2-form (equivalently the Riemann tensor) for the more general ansatz (\ref{expansatz}) above,
\begin{equation}
R^{0i}=H_i(H_i-H_0)e^{-2H_0t}e^0\wedge e^i \quad , \qquad R^{ij}=H_iH_j e^{-2H_0t}e^i \wedge e^j~,
\end{equation}
and notice that for every value $H_0\neq 0$ the Riemann components diverge either to the past or to the future (at the singularity) in which case the dominant term in the Lovelock equation will correspond again to pure Lovelock. The exponential case, $H_0=0$, is quite different as the components of the Riemann tensor become constant and all Lovelock terms contribute to the same order in the equation of motion. Besides, every curvature scalar will be constant 
and there is no curvature singularity at all. The geometry is completely regular.

In the case of Einstein gravity we can add a positive cosmological constant to get solutions of this type. In the general case we would get a complicated set of polynomial constraints involving all of the Lovelock couplings. Even in the pure Lovelock case, the addition of just a cosmological constant term would complicate the equations to a high degree. 

In analogy with the Kasner metrics we can also discuss the values the different curvature tensors take for vacuum exponential solutions. We summarize the situation in Table \ref{classExp}. The only difference is that the subset of sector {\bf (b)} for which $\mathcal{R}=0$ disappears. This is easy to understand as it would correspond to $H_i=H_0$ in the general parametrization, thus $H_i=0$ and they reduce to type {\bf (c)} metrics. For this class of solutions we have a perfect hierarchy of solutions with regard to which curvature tensors vanish. This is true for $N=2$ but expected to hold in general. Type {\bf (a)} vacua have $R=\mathcal{R}=\mathbb{R}=0$, type {\bf (c)} $\mathcal{R}=\mathbb{R}=0$, $R\neq 0$, type {\bf (b.1)} (or {\bf(b)} for d=2N+1) has $\mathbb{R}=0$, $R,\mathcal{R}\neq0$  and finally for type {\bf (b.2)} all those tensors are non-zero. The Riemann tensor is only zero for the trivial Minkowski metric in this case --there is no analogue of {\it flat Kasner}. We can then distinguish the different families of vacuum solutions just by using these tensors!

In the case of Kasner metrics this is not true due to the existence of exceptions; {\it i.e.} {\it flat Kasner}, that belongs to type {\bf (c)} yet has $R=0$, and the subset of type {\bf (b)} metrics with all non-zero exponents equal to one, that have $\mathcal{R}=0$. Once this cases are taken care of separately, the above classification scheme carries over to Kasner as well.

\begin{table}[ht]
\begin{tabular}{c|c||c|c||c}
type & $d$ & isotropy cond. & vacuum & vac. curvature\\
\hline\hline
{\bf (a)} & any $d$ & $H_i=H \ ,\quad \forall i=1,2\ldots d-1$ & $H=0$ & $R=\mathcal{R}=\mathbb{R}=0$\\
\hline
{\bf (b)} & $2N+1$ &  & $H_1=0$ & $\mathbb{R}=0$, $R,\mathcal{R}\neq0$\\
\hskip.4in {\bf --b.1--}\, & $2N+2$  & $\sum_{i=1}^{d-1}H_i=0$ & $H_1=H_2=0$ & $\mathbb{R}=0$, $R,\mathcal{R}\neq0$\\
\hskip.4in {\bf --b.2--}\, & $2N+2$  &  & $\sum_{i=1}^{d-1}\frac1{H_i}=0$ & $ R,\mathcal{R},\mathbb{R}\neq 0$\\
\hline
{\bf (c)} & $2N+1$ & $H_1=H_2=0$ & \multirow{2}{*}{\it all} & \multirow{2}{*}{$\mathcal{R}=\mathbb{R}=0$, $R\neq0$}\\
& $2N+2$ & $H_1=H_2=H_3=0$ & & \\
\hline
{\bf (d)} & \multirow{2}{*}{$2N+2$} & $H_i=H_{1,2}$ with multiplicities $n_{1,2}$ & \multirow{2}{*}{\it none} & \\
&  & $\frac{n_1-1}{H_1}+\frac{n_2-1}{H_2}=0~, \quad n_1+n_2=2N+1$ & &
\end{tabular}
\caption{Classification of isotropic and vacuum exponential type solutions in pure Lovelock gravity.}
\label{classExp}
\end{table}

\section{Discussion}
Kasner solutions play a fundamental role in the analysis of big bang singularities. 
In this note we have analyzed and classified Kasner type metrics in pure Lovelock theories of gravity. Regardless of the phenomenological interest of such theories, they capture the leading order dynamics in the approach to the singularity within the Lovelock class of theories. Close to the big-bang the curvature is generically diverging, thus the leading contribution will be that of the highest order curvature term, thus pure $N$th order Lovelock gravity in $d=2N+1,2N+2$.

Analyzing the conditions for isotropy we were able to classify perfect fluid and vacuum solutions in several families denoted as {\bf (a)}-{\bf(d)} (just {\bf (a)}-{\bf(c)} in the vacuum case, with subtypes {\bf (b.1)} and {\bf (b.2)} in even dimensions). In vacuum the different types differ mainly in the number of flat directions, $p_i=0$, the metric has. Type {\bf (a)} vacuum metrics have all directions flat, it is just Minkowski. Type {\bf (c)} metrics have at least $d-2N+1$ flat directions, thus two and three in odd and even dimensions respectively, whereas type {\bf (b)} solutions have at most $d-2N$ vanishing $p_i$. In odd dimensions these have one flat direction but in even dimensions they may have either two or none, corresponding to subtypes {\bf (b.1)} and {\bf (b.2)}. Lovelock non-flat vacua emerge only in even $d=2N+2$ dimensions only when all exponents $p_i$s are non-zero, these satisfying the conditions $\sum p_i = 2N-1$  and $\sum p_i^{-1}=0$. This is true for any pure Lovelock theory in vacuum.

Parallely, we analyzed the values the different Lovelock-Riemann analogue tensors take for the different vacuum types. We found a nice correspondence between the different families of solutions and the vanishing of particular sets of these 4th rank tensors, in a hierarchical way. Type {\bf (a)} solutions verify $R=\mathcal{R}=\mathbb{R}=0$, type {\bf (c)} have $\mathcal{R}=\mathbb{R}=0$ and $R\neq 0$ (except for {\it flat Kasner} that has $R=0$), while type {\bf (b.1)} --or just {\bf (b)} in odd dimensions-- yields $\mathbb{R}=0$ and $R,\mathcal{R}\neq0$ (except for the solutions with all non-zero exponents equal to unity). Finally type {\bf (b.2)} solutions are the ones for which all these tensors are non-vanishing. We can then use this tensors to classify our solutions in the different families, taking into account the exceptions. This may be very helpful in classifying solutions in case these are not given in the canonical Kasner form, but in some other possibly complicated set of coordinates. Besides, this classification scheme may be relevant for more general classes of solutions. 

A similar classification scheme exists for another class of solutions closely related to Kasner. Exponential type metrics as those studied in section VII, are also divided in isotropy types as Kasner's and the relation to the vanishing of the different sets of tensors carries over, in this case without exceptions. The {\bf (b.2), (b.1), (c), (a)} types can be defined precisely setting to zero one further 4-tensor at a time; $\mathbb{R}=0$, $\mathcal{R}=\mathbb{R}=0$, $R=\mathcal{R}=\mathbb{R}=0$.


In four dimensional Einstein gravity ($N=1$ pure Lovelock) we can analyze much more general cosmological models. These have been classified long ago by Bianchi. Kasner metrics correspond to the Bianchi type I models in this classification, and represent the asymptotic solutions within those models. The situation is much more complicated for more general models, yet Kasner solutions still play a preeminent role. Bianchi type II models turn out to have Kasner asymptotics both to the past and the future. The Kasner solutions connected through a Bianchi type II trajectory define the so called {\it Kasner map}. For more general models, the BKL conjecture proposed that the Universe close to the initial singularity undergoes a series of oscillations, transitions between Kasner epochs where the expanding and contracting directions exchange their roles. A precise realization of this conjecture is given by the {\it Mixmaster Universe}. This represents the asymptotic behavior of Bianchi types VII and XI and is given by the iteration of the Kasner map, giving in this way the sequence of Kasner epochs. This sequence is infinite in four dimensions and the dynamics has been shown to be chaotic \cite{Cornish1997}. This chaotic behavior disappears in higher dimensional Einstein gravity.

Non-trivial Kasner metrics in four dimensions correspond to type {\bf (b)} solutions in our classification (type {\bf (b.1)} are just {\it flat Kasner} in that case). It is reasonable to expect that a similar behavior to that of the Mixmaster attractor may also appear for other even dimensional pure Lovelock theories. It would be one more feature of $d=4$ Einstein gravity respected by pure Lovelock theories in all dimensions. Our classification could also be regarded as the necessary first step in such more general analysis. The obvious next step would be trying to generalize a result analogous to that of Bianchi type II in four dimensions and define a {\it generalized Kasner map}. For that we would have to introduce curvature in the spatial slices of our model (equivalently a nontrivial Lie group structure). The analogue of Bianchi II models would be a deformation in just a given 3-dimensional subspace; {\it i.e.} $t^{p_i}dx_i\to e^i$ with $i=1,2\ldots d-1$ and
$$ [e^1,e^2]=n(t) e^3 $$
in the usual notation, the rest of the commutation relations being zero. This is the simplest possible modification of the Lie group structure. If a structure such as that of the Kasner map exists also in this case, this could also provide new examples of chaotic maps, interesting objects in their own right. The rich geometric structure we have found in the space of type {\bf (b)} solutions would perhaps  come in handy in performing such analysis. 



\appendix*
\section{Closed timelike curves (CTCs) from complex Kasner exponents}
\label{app}

The Kasner type metrics that are the subject of this paper are solutions of pure Lovelock gravity independently of the signature considered. The discussion of the next paragraphs will involve just the form of the metric and not the details of the equations, in that respect all that is said in this section applies also to Einstein gravity on any dimension. The choice of signature used in the previous sections corresponds to the description of the near region to a spacelike singularity. We can also describe timelike singularities in a similar fashion, we just have to exchange the roles of the time and one of the spacelike coordinates, $t\leftrightarrow x$. The form of the solutions, {\it i.e.} the possible values of the exponents is independent of the choice of signature, at least for real values of these exponents.

Formally one can also choose complex values for the $p_i$, as soon as they verify the relevant algebraic equations. The simplest such case is that of a complex conjugate pair of exponents, the rest being real. This is also an option in 4d Einstein gravity for which we can solve for two of the exponents in terms of the third as
\begin{equation}
p_\pm=\frac{1-p_3\pm\sqrt{(1-p_3)(3p_3+1)}}{2}
\end{equation}
These are real for a small range of values of $p_3\in [-1/3,1]$, otherwise they have an imaginary part and we can write $p_\pm=a+b\,i$, for real $a,b$. Notice that this is the range of values considered in the Lifshitz-Khalatnikov parametrization,
\begin{equation}
p_1=\frac{-u}{1+u+u^2}\quad ; \qquad p_2=\frac{1+u}{1+u+u^2}\quad ; \qquad p_3=\frac{u(1+u)}{1+u+u^2}
\end{equation}
with real parameter $u$. Even though $u$ may take any real value the exponents always belong to the range $p_i\in [-1/3,1]$.

\begin{figure}[b]
\begin{center}
\includegraphics[scale=.7]{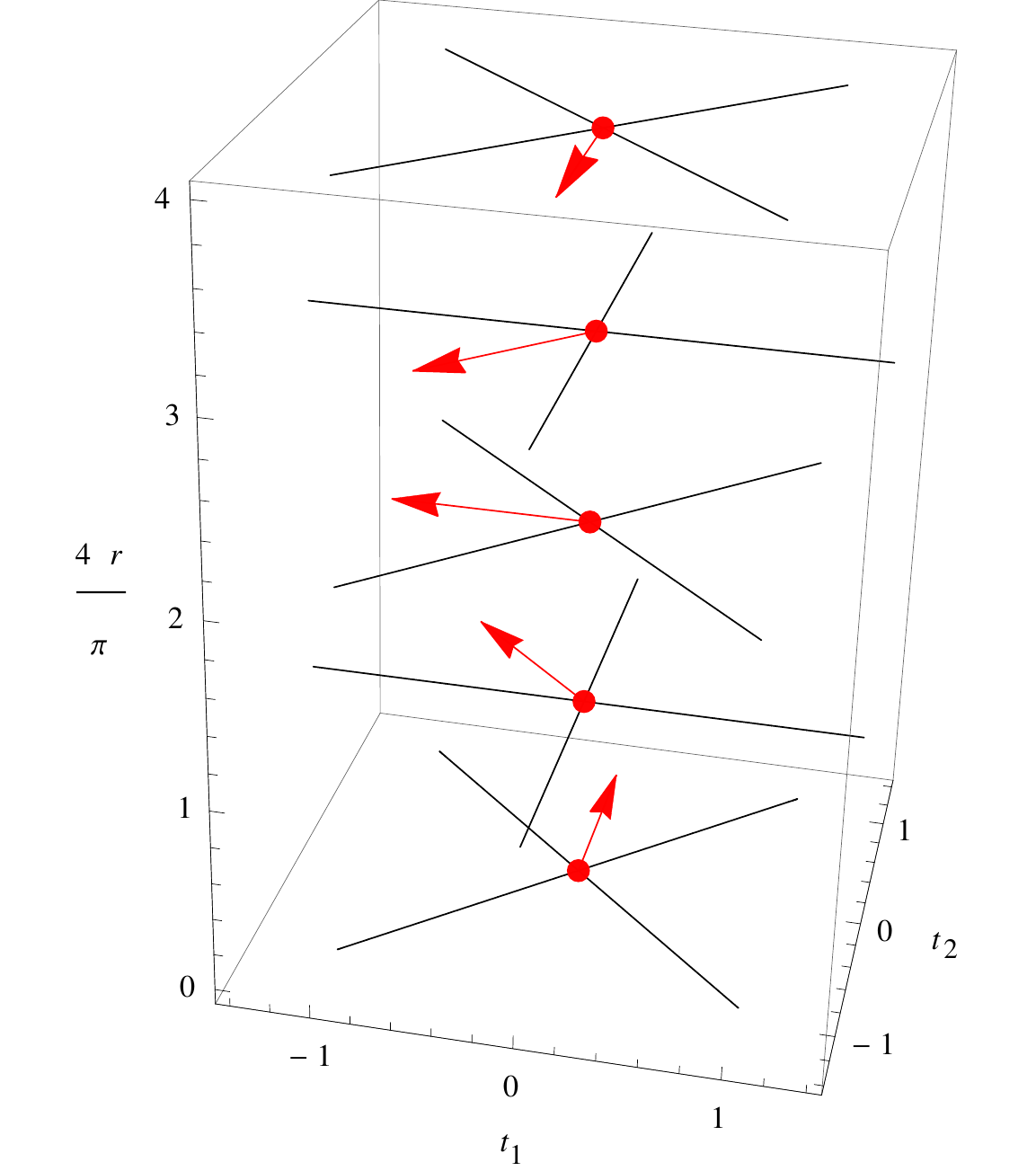}~\includegraphics[scale=.8]{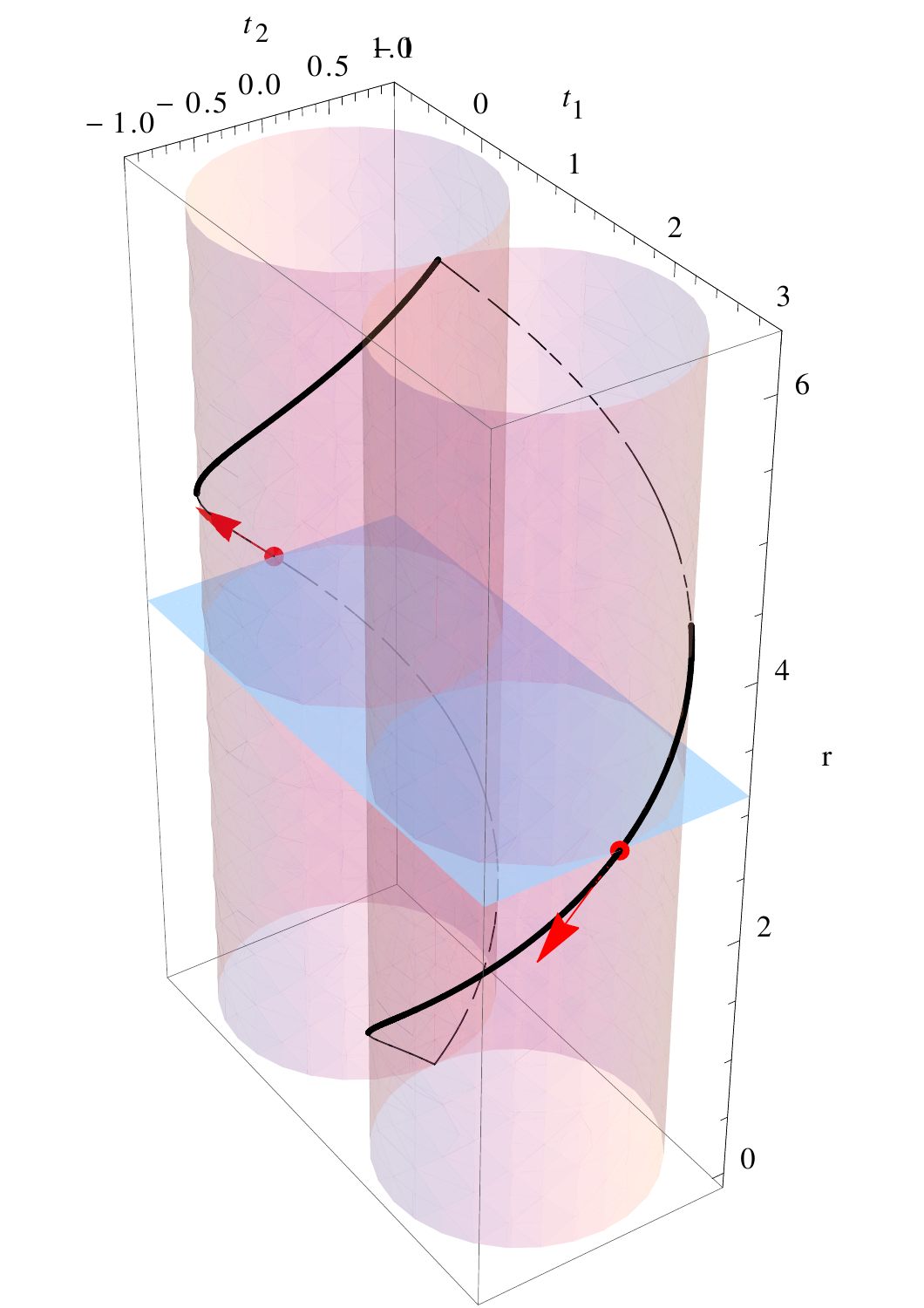}
\end{center}
\caption{Causal structure (on the left) and close timelike curves (on the right) on the $(r,t_1,t_2)$ space of metric (\ref{complexK}).}
\label{CTC}
\end{figure}

Starting with Riemannian signature with a complex Kasner metric of the form,
\begin{equation}
ds^2=e^{2r}dr^2+\sum_{i\neq 1,2} e^{2p_i r}dx_i^2 + e^{2ar}\left(e^{2ibr}dx_1^2+e^{-2ibr}dx_2^2\right)~,
\label{complexK}
\end{equation}
the term in brackets of the above metric is complex, this being the reason complex Kasner exponents have been disregarded in the past. We can however bring the metric to a real (Lorentzian) form performing a {\it complex} change of coordinates, $x_1=z=t_1-i\,t_2;\ x_2=\bar{z}=t_1+i\,t_2$. This yields,
\begin{equation}
ds^2=e^{2r}dr^2+\sum_{i\neq 1,2}e^{2p_i r}dx_i^2+2e^{2ar}\left(\cos(2br)(dt_1^2-dt_2^2)+2\sin(2br)\,dt_1\,dt_2\right)
\end{equation}
even though we started off with Riemannian signature the presence of complex exponents already implies one timelike direction when the metric is written in real form. Notice however that the direction of the timelike direction in the $(t_1,t_2)$ plane changes as we move along the radial direction $r$ (see Figure \ref{CTC}, left). This makes a bit difficult to visualize the causal structure of this space and will determine some of its peculiar properties, namely the existence of closed timelike curves. Have we considered more than two complex exponents we would end up with  more than one timelike direction. In some situations it is not even possible to bring the metric to any real form.

For simplicity we will analyze the case $a=1$ in detail. For $a\neq 1$ some of the expressions will be a bit more complicated but the qualitative picture is still the same. We will consider just the $(r,t_1,t_2)$ part of the metric and neglect the conformal factor $e^{2ar}$ as it does not change the causal structure. Equivalently, we consider
\begin{equation}
d\hat{s}^2=dr^2+\cos(2br)(dt_1^2-dt_2^2)+2\sin(2br)\,dt_1\,dt_2 ~.
\end{equation}
We fix the future direction as that of positive $t_2$ at $r=0$. This fixes the future direction for the rest of the space. It is orientable. For any $r=\text{const.}$ plane it is easy to find the null lines as
\begin{equation}
t_1=\pm\sqrt{\frac{1\mp \sin(2br)}{1\pm \sin(2br)}}t_2
\end{equation}
These are $t_1=\pm t_2$ for $r=0$ and they rotate counterclockwise with constant opening as we increase $r$ (see Figure \ref{CTC}).
In addition to these, there is another family of null trajectories that is particularly simple to find. That is
\begin{equation}
t_1(r)=\frac1b\cos(br)\quad ;\qquad t_2(r)=\frac1b\sin(br)
\end{equation}
that is future directed for $\dot{r}>0$. We can also have a future directed null trajectory going to lower  values of $r$, {\it i.e.} $\dot{r}<0$, just changing the overall sign of $t_i$,
\begin{equation}
t_1(r)=\frac1b(2-\cos(br))\quad ;\qquad t_2(r)=-\frac1b\sin(br)
\end{equation}
where we added a constant such that both curves pass by $(\frac1b,0,0)$. These two null curves coincide also for $r=2n\pi/b$ but they pass by these points in reversed order. The first curve is future directed from $r=0$ to $r=2\pi/b$ whereas the second is future directed from $r=2\pi/b$ to $r=0$. We can then take these two pieces and construct a closed future directed null curve (see Figure \ref{CTC}, right). This is the simplest case of such a curve, it is easy to work out many other examples. In the same way it is possible to demonstrate the existence of closed timelike curves, the null case is just some limiting case that is easier to analyze. In fact it is enough to add some small timelike pieces in between the trajectories going in both directions. This is true independently of the values of the complex exponents (equivalently $a$ and $b$).

\end{document}